\documentclass[%
 reprint,
 amsmath,amssymb,
 aps,
pra,
]{revtex4-1}

\usepackage[dvipdfmx]{graphicx}
\usepackage{dcolumn}
\usepackage{bm}

\usepackage{color}

\begin{document}


\title{Phase diagram of vortices in the polar phase of spin-1 Bose--Einstein condensates
 }

\author{Hiromitsu Takeuchi}
\email{takeuchi@osaka-cu.ac.jp}
\homepage{http://hiromitsu-takeuchi.appspot.com}
\affiliation{
Department of Physics and Nambu Yoichiro Institute of Theoretical and Experimental Physics (NITEP),\\
 Osaka City University, Osaka 558-8585, Japan
}




\date{\today}

\begin{abstract}
The phase diagram of lowest-energy vortices in the polar phase of spin-1 Bose--Einstein condensates is investigated theoretically.
Singly quantized vortices are categorized by the local ordered state in the vortex core and
three types of vortices are found as lowest-energy vortices, which are elliptic AF-core vortices, axisymmetric F-core vortices, and N-core vortices.
These vortices are named after the local ordered state, ferromagnetic (F), antiferromagnetic (AF), broken axisymmetry (BA), and normal (N) states apart from the bulk polar (P) state.
The N-core vortex is a conventional vortex, in the core of which the superfluid order parameter vanishes.
The other two types of vortices are stabilized when the quadratic Zeeman energy is smaller than a critical value.
The axisymmetric F-core vortex is the lowest-energy vortex for ferromagnetic interaction, and it has an F core surrounded by a BA skin that forms a ferromagnetic-spin texture, as exemplified by the localized Mermin--Ho texture.
The elliptic AF-core vortex is stabilized for antiferromagnetic interaction; the vortex core has both nematic-spin and ferromagnetic orders locally and is composed of the AF-core soliton spanned between two BA edges.
The phase transition from the N-core vortex to the other two vortices is continuous, whereas that between the AF-core and F-core vortices is discontinuous.
The critical point of the continuous vortex-core transition is computed by the perturbation analysis of the Bogoliubov theory and the Ginzburg--Landau formalism describes the critical behavior.
The influence of trapping potential on the core structure is also investigated.
\end{abstract}

\pacs{Valid PACS appear here}
\maketitle

\section{Introduction}

In an ordered state after spontaneous symmetry-breaking phase transition,
 the ground states of the considered system are energetically degenerate.
 The ordered state is described by the order parameter field and
the topology of the order parameter space depends on the type of symmetry that is broken via the phase transition.
In multicomponent superfluids,
topological defects can take a variety of structures according to the multidegree of freedom of the order parameters,
 such as domain walls (solitons), vortices (strings), and monopoles (hedgehogs) as two-dimensional, one-dimensional, and zero-dimensional defects in three dimensions  \cite{vollhardt2013superfluid,volovik2003universe,kasamatsu2005vortices,kawaguchi2012spinor}.
The type of defect formed depends on the symmetry of the order parameter space of the ground (bulk) state.

 Recently, it has been shown theoretically that a singly quantized vortex can have a nonaxisymmetric form in the polar (P) phase of a spin-1 Bose--Einstein condensate (BEC) with a quadratic Zeeman shift \cite{takeuchi2020quantum}.
A nonaxisymmetric vortex, called an elliptic vortex, is considered the equilibrium state of a wall--vortex composite defect observed in an experiment of a spin-1 BEC \cite{kangPhysRevLett.122.095301} while the composite defect was thought to be dynamically unstable owing to the snake instability.
The elliptic vortex is stabilized by the appearance of a local ordered state in the vortex core with a symmetry different from that in the bulk state.
An elliptic vortex is considered as the Joukowski transform of an axisymmetric vortex, and thus its hydrodynamic behavior is different from that of a conventional axisymmetric vortex.
These facts strongly suggest that not only the symmetry of the bulk ordered state but also the local ordered state in the core of topological defects may be crucial to the properties of topological defects in multicomponent superfluids.

The classification of topological defects in spinor BECs has been performed extensively based on the homotopy theory by considering the symmetry and phase diagram of the bulk ordered state with respect to the interatomic interaction and the Zeeman shift (see Ref.~\cite{kawaguchi2012spinor} and the references therein).
However, few systematic studies have investigated which type of local ordered state is energetically preferred in the core of topological defects by taking into account the Zeeman shift.
Recently, the impact of the quadratic Zeeman shift on the core of solitons and vortices has been investigated systematically \cite{liu2020phase,underwood2020properties}
 and it has been shown that the size of topological defects can diverge in the zero limit of the quadratic Zeeman shift in the P and antiferromagnetic (AF) phases of spin-1 BECs.
These results imply that the property of topological defects can be sensitive to the finite-size effect or system boundary when the quadratic Zeeman shift is small.
In other words, the fundamental property in uniform systems can be ambiguous in a trapping system when the condensate size or the Thomas-Fermi radius is comparable with the characteristic size of a defect determined by the quadratic Zeeman shift, called the Zeeman length.
Nevertheless the core structure of solitons in trapped spin-1 BECs with zero/non-zero quadratic Zeeman shift has been investigated in the literature without recognizing the importance of the finite-size effect characterized by the Zeeman length \cite{isoshima2001quantum,PhysRevA.93.033633,katsimiga2020phase}.
In Ref.~\cite{liu2020phase} the core structure of a soliton in uniform systems and its phase diagram in the P phase have been revealed in the context of spontaneous symmetry breaking, leading to the prediction of the F-core soliton observed in the recent experiment \cite{PhysRevLett.125.170401}.
On the other hand, the phase diagram of vortices in the P phase has never been investigated in a proper manner under the effect of the quadratic Zeeman shift,
although it has been done partly for a $^{23}$Na condensate with antiferromagnetic spin interaction \cite{takeuchi2020quantum}.
It should be also mentioned that there are a few researches on the fundamental aspect of vortex dynamics under the quadratic Zeeman effect \cite{PhysRevLett.103.080603,PhysRevA.94.063615,williamson2020damped}.
This is in contrast to the fact that a good understanding of a rich variety of topological defects in superfluid $^3$He has been developed with experimental and theoretical investigations of the core structure \cite{vollhardt2013superfluid,volovik2003universe}, such as the phase diagram of vortices in the A and B phases of superfluid $^3$He \cite{PhysRevLett.75.3320,PhysRevB.101.024517}.

\begin{figure}
\begin{center}
\includegraphics[width=1.0 \linewidth, keepaspectratio]{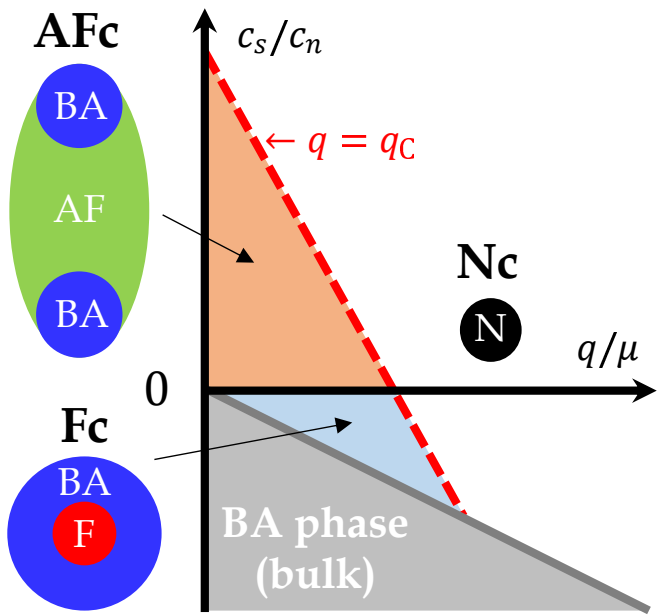}
\end{center}
\vspace{-5mm}
\caption{
Phase diagram of the lowest-energy vortices and schematic of three types of singly quantized vortices in the P phase of spin-1 BECs.
An elliptic AF-core vortex (AFc) has two BA edges with opposite transverse spin density, and the local F state around the vortex axis with longitudinal spin density is surrounded by a BA skin with transverse spin density in an axisymmetric F-core vortex (Fc) (see also Fig.~\ref{Fig-Cross}).
The conventional vortex, called the N-core vortex (Nc) in this paper, is also described for reference.
The phase transition at the boundary between the regions of the N-core vortex and the AF- or F-core vortex in the phase diagram is continuous,
whereas the boundary $c_s=0$ between the AF- and F-core vortices corresponds to a discontinuous phase transition.
The former boundary is described by $\frac{q_{\rm C}}{\mu}=-(1+\tilde{M})\frac{c_s}{c_n}-\tilde{\epsilon}$ [Eq.~(\ref{eq:qc_mu})] with dimensionless constants $\tilde{M}\approx 0.45$ and $\tilde{\epsilon}=-0.25$.
The phase boundary of the BA phase in the bulk (grey line) is given by $\frac{c_s}{c_n}=-\frac{q}{2\mu}$.
}
\label{Fig-diagram}
\end{figure}

In this work, the core structure of a singly quantized vortex in the P phase of spin-1 BECs in uniform systems is investigated theoretically by changing the parameters associated with the spin interaction and the quadratic Zeeman energy without external rotation.
It is shown that the ordered states other than the bulk state, the AF, broken axisymmetry (BA), and ferromagnetic (F) states, occur locally in the vortex core with different configurations depending on the parameters.
The theoretical predictions are summarized as the phase diagram of vortices and the schematic vortex-core structures in Fig.~\ref{Fig-diagram}.
These predictions are consistent with numerical analyses and can be examined by experiments on spin-1 BECs.

The reminder of this paper is organized as follows.
In Sec.~\ref{sec-F}, we briefly introduce the background theory to understand the main contents of this work by focusing on the aspect of the nematic-spin order.
In Sec.~\ref{sec-N}, we provide a brief overview of the phase diagram of vortices and demonstrate the numerical result by showing the typical vortex structure for antiferromagnetic and ferromagnetic interaction in Fig.~\ref{Fig-Cross}.
Detailed analyses are described in the following sections and thus
readers who want to understand the main contents quickly can skip to Sec.~\ref{sec-S}, after reading Sec.~\ref{sec-N}.
Section~\ref{sec-AV} is mainly devoted to a theoretical description of the physical interpretation of the vortex-core structure in an axisymmetric vortex based on the hydrostatic approximation.
This approximation provides a systematic method for us to determine the local ordered states in the vortex core.
In Sec.~\ref{sec-EV},
the structure of an elliptic vortex is explained from a different viewpoint from that in Ref.~\cite{takeuchi2020quantum}.
In Sec.~\ref{sec-TN}, the continuous phase transition of a normal-core vortex is described by the perturbation theory of the Bogoliubov theory, and the Ginzburg--Landau formalism is introduced to describe the critical behavior of the vortex-core phase transition.
The influence of the trapping potential and the finite-size effect is mentioned from a general perspective in the context of the hydrostatic approximation.
We conclude in Sec.~\ref{sec-S}, with a summary and discussions of the relation to similar problems in other systems and the future prospects of this work.

\section{Formulation}\label{sec-F}
Here, the formulation of the concept of the nematic-spin order and certain fundamental energetics are introduced.
We restrict the contents to the minimum necessary to understand the main part of this work.
Readers may refer to the review paper \cite{kawaguchi2012spinor}, for example, for full details of the conventional formulation and the phase diagram of spin-1 BECs.

\subsection{Lagrangian}

Spin-1 BECs are described by the macroscopic wave functions $\Psi_m~(m=0,\pm 1)$ of the $|m \rangle$ Zeeman component at low temperatures.
To express the vortex states in the P phase, it is convenient to introduce the Cartesian representation \cite{ohmi1998bose}
\begin{eqnarray}
{\bm \Psi}=
\left[
\begin{array}{c}
\Psi_x \\
\Psi_y \\
\Psi_z \\
\end{array}
\right]
=\left[
\begin{array}{c}
\frac{-1}{\sqrt{2}}(\Psi_{+1}-\Psi_{-1}) \\
\frac{-i}{\sqrt{2}}(\Psi_{+1}+\Psi_{-1}) \\
\Psi_0 \\
\end{array}
\right].
  \label{eq:Cartesian}
\end{eqnarray}
This system obeys the Lagrangian in the Gross--Pitaevskii (GP) model,
\begin{eqnarray}
  {\cal L}({\bm \Psi})=\int d^3x i\hbar {\bm \Psi}^*\cdot \partial_t {\bm \Psi}-G({\bm \Psi}),
\label{eq:L}
\end{eqnarray}
with the thermodynamic energy functional
\begin{eqnarray}
  G=\int d^3x\left[\frac{\hbar^{2}}{2M}\sum_{j=x,y,z}(\partial_j{\bm \Psi}^*)\cdot(\partial_j{\bm \Psi})+ {\cal U}\right]
\label{eq:G}
\end{eqnarray}
and
\begin{eqnarray}
{\cal U}=\frac{c_n}{2}n^2
+\frac{c_s}{2}{\bm s}^2
 - q \left| \Psi_z\right|^2 -(\mu-q) n -ps_z.
\label{eq:energy_density}
\end{eqnarray}
Here, the condensate density and the spin density are represented by the dot product
\begin{eqnarray}
  n={\bm \Psi}^*\cdot{\bm \Psi}=\sum_m \left| \Psi_m \right|^2
\end{eqnarray}
and the cross product
\begin{eqnarray}
  {\bm s}=
  \left[
  \begin{array}{c}
  s_x \\
  s_y \\
  s_z \\
  \end{array}
  \right]
  =i {\bm \Psi}\times {\bm \Psi}^*
  =\left[
  \begin{array}{c}
  \sqrt{2}{\rm Re}\left[(\Psi_{+1}+\Psi_{-1})\Psi_0^*\right] \\
  \sqrt{2}{\rm Im}\left[\Psi_0(\Psi_{+1}^*-\Psi_{-1}^*) \right] \\
  |\Psi_{+1}|^2-|\Psi_{-1}|^2 \\
  \end{array}
  \right],
\end{eqnarray}
respectively.
The coupling constants $c_n$ and $c_s$ are expressed in terms of particle mass $M$ and $s$-wave scattering length $a_F$ of the total spin-$F$ channel as $c_n=\frac{4\pi \hbar^{2}}{3M}(2a_2+a_0)$ and $c_s=\frac{4\pi \hbar^{2}}{3M}(a_2-a_0)$.
The linear and quadratic Zeeman shifts are parametrized by $p$ and $q$, respectively.
The chemical potential $\mu$ and $p$ are the Lagrange multipliers associated with the conservation of the total particle number $\int d^3x n$ and the total magnetization $\int d^3x s_z$ along the spin quantization axis.
Here, we consider the ``nonbiased'' case of $p=0$, as has been realized experimentally \cite{kangPhysRevLett.122.095301,kangPhysRevA.101.023613}.

\subsection{Nematic-spin order parameter}

In spin-1 BECs with $p=0$, there are four ordered phases with different ground states, which are the
P, ferromagnetic (F), antiferromagnetic (AF), and broken axisymmetry (BA) phases.
The ground state in uniform systems is obtained by minimizing ${\cal U}$ with the four given variables ($\mu$, $q$, $c_n$, and $c_s$) fixed.
We consider ordered states with finite amplitude of order parameters for $\mu>0$ and $c_n>0$
whereas the normal (N) state ${\bm \Psi}=0$ is the ground state for $\mu \leq 0$ with $c_n>0$.
To express the different ordered states in a unified manner,
the order parameter is represented as
\begin{eqnarray}
   {\bm \Psi}=e^{i\Theta}({\bm d}+i{\bm e})
   \label{eq:evec}
\end{eqnarray}
with the global phase $\Theta$ and the real vectors ${\bm d}=[d_x,d_y,d_z]^{\rm T}$ and ${\bm e}=[e_x,e_y,0]^{\rm T}$.
The spin density is written as ${\bm s}=2{\bm d}\times {\bm e}$.

For the ground states of zero spin density ${\bm s}=0$ with ${\bm e}=0$ in the P and AF phases,
 the order parameter reduces to
\begin{eqnarray}
   {\bm \Psi}=\sqrt{n}e^{i\Theta}\hat{\bm d}
   \label{eq:evec2}
\end{eqnarray}
with the unit vector $\hat{\bm d}$.
The ground states in the P and AF phases correspond to $\hat{\bm d}\cdot\hat{\bm z}=\pm 1$ and $\hat{\bm d}\cdot\hat{\bm z}=0$, respectively.
The vector $\hat{\bm d}$ is called the pseudo-director because its behavior is similar to the director $\tilde{\bm d}$ in a uniaxial nematic liquid crystal \cite{chandrasekhar_1992}.
In liquid crystals, the ordered state of $\tilde{\bm d}$ is identical to that of $-\tilde{\bm d}$.
The nematic behavior is imitated by combining $\hat{\bm d}$ with the global phase $\Theta$;
the ordered state is invariant under the operation $(\hat{\bm d},\Theta)\to (-\hat{\bm d},\Theta+\pi)$.
This property is sometimes called the nematic-spin or spin-nematic order.

To explicitly reveal the condition ${\bm s}=0$ for the nematic-spin order,
we write the square of the spin density as ${\bm s}^2=s_\bot^2+s_z^2$ with
\begin{eqnarray}
s_\bot^2=2|\Psi_{0}|^2\left( |\Psi_{+1}|^2+|\Psi_{-1}|^2+2|\Psi_{+1}||\Psi_{-1}|\cos\delta\Theta \right)
   \label{eq:sbot}
\end{eqnarray}
and
\begin{eqnarray}
\delta\Theta =\arg\Psi_{+1}+\arg\Psi_{+1}-2\arg\Psi_{0}.
   \label{eq:dTheta}
\end{eqnarray}
The nematic-spin order is realized when $\delta\Theta=\pi$ and $s_z=0$ with $|\Psi_{+1}|=|\Psi_{-1}|$.

The P phase, our target phase, occurs for $q>0$ and $\frac{c_s}{c_n}>-\frac{q}{2\mu}$,
and the ground state is the P state ${\bm \Psi}={\bm \Phi}_{\rm P}$,
\begin{eqnarray}
  {\bm \Phi}_{\rm P}=\pm \sqrt{n_{\rm P}} e^{i\Theta}\hat{\rm z}
   \label{eq:P}
\end{eqnarray}
with $n_{\rm P}=\frac{\mu}{c_n}$.
In terms of $\Psi_m$, the P state is $(\Psi_0,\Psi_{\pm 1})=(\sqrt{n_P}e^{i\Theta_0},0)$.
The ground state in the AF phase ($q <0$ and $c_s > 0$) is written as ${\bm \Phi}_{\rm AF}=\sqrt{n_{\rm AF}} e^{i\Theta}\hat{\bm \rho}$ with $\hat{\bm \rho}\bot \hat{\bm z}$.
In the BA and F phases, ${\bm s}$ is parallel and perpendicular to the $z$ axis with ${\bm s}=\pm \hat{\bm z}$ and ${\bm s}=\hat{\bm \rho}$, respectively.

\begin{figure*}
\begin{center}
\includegraphics[width=1.0 \linewidth, keepaspectratio]{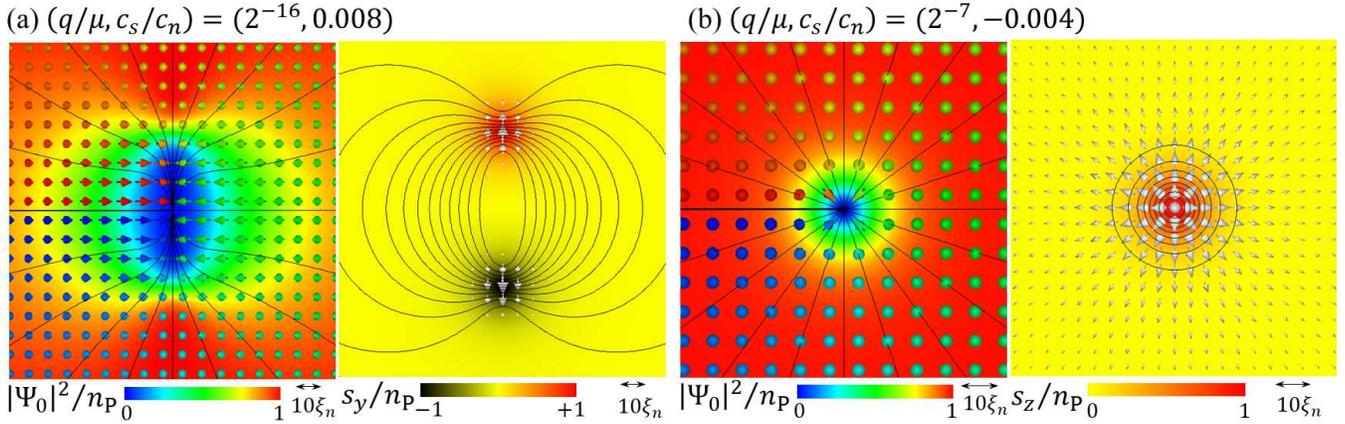}
\end{center}
\vspace{-5mm}
\caption{
The cross-section profiles of $|\Psi_0|^2$ (left) and ${\bm s}$ (right) in the core of (a) an AF-core vortex with $(\frac{q}{\mu},\frac{c_s}{c_n})=(2^{-16},0.008)$ and (b) an F-core vortex with $(2^{-7},-0.004)$.
Left: The direction and color of the arrows represent the unit vector $\frac{{\bm d}}{|{\bm d}|}$ and phase $\Theta_0=\arg\Psi_0$, respectively. The black curves show the contour of $\Theta_0$.
The nematic-spin order is demonstrated as the texture of the pseudo-director field $\hat{\bm d}=\frac{{\bm d}}{|{\bm d}|}$ for ${\bm s}=0$,
and $\hat{\bm d}$ is ill-defined for ${\bm s}\neq 0$.
Right: The direction and length of the arrows imply those of the spin density ${\bm s}$.
The contour of $|\Psi_{+1}|^2$ is represented by the black curves. We have $|\Psi_{+1}|^2=|\Psi_{-1}|^2$ for the AF-core vortex, whereas $|\Psi_{-1}|^2$ is negligibly small for the F-core vortex.
The length scale of each plot is shown by a double-headed arrow.
}
\label{Fig-Cross}
\end{figure*}

\subsection{Vortex energy}
The energy of a straight vortex is defined as the excess energy in the presence of the vortex in a cylindrical system.
The vortex energy per unit length is written as
\begin{eqnarray}
  E_{\rm vortex}=\frac{G({\bm \Phi}_{\rm vortex})-G({\bm \Phi}_{\rm bulk})}{L_z},
   \label{eq:Evor}
\end{eqnarray}
where ${\bm \Psi}={\bm \Phi}_{\rm vortex}$ is a stationary state of a straight vortex, ${\bm \Phi}_{\rm bulk}={\bm \Phi}_{\rm P}$ is the bulk state, and $L_z$ is the length of the vortex along the $z$ axis.
The value of Eq.~(\ref{eq:Evor}) represents the tension of the vortex, which is important for determining its dynamical properties, such as Kelvin waves and vortex rings \cite{donnelly1991quantized}.
In this work, we focus on the static properties, especially the internal structure of the vortex core.

\subsection{Characteristic lengths}
For $p=0$, the P phase is parametrized by two dimensionless variables, $\frac{q}{\mu}$ and $\frac{c_s}{c_n}$
 by rescaling energy and length by $\mu$ and the density healing length
\begin{eqnarray}
  \xi_n=\frac{\hbar}{\sqrt{M\mu}},
   \label{eq:xi_n}
\end{eqnarray}
 respectively.
In our system, the quadratic Zeeman shift $q$ gives another important scale, called the Zeeman length,
\begin{eqnarray}
  \xi_q=\frac{\hbar}{\sqrt{Mq}}.
   \label{eq:xi_q}
\end{eqnarray}
This length is introduced to characterize the core size of the AF-core soliton in the P phase \cite{liu2020phase} and the nematic-spin vortex in the AF phase \cite{underwood2020properties},
where the local phase transition in the core is determined by the competition between the density healing length and the Zeeman length.

\section{Phase diagram}\label{sec-N}

The phase diagram of vortices is summarized in Fig.~\ref{Fig-diagram}.
Here, we give a qualitative interpretation of the phase diagram mainly based on the facts from the numerical results.
The detailed theoretical analysis is demonstrated in the following sections.

The core structure of a singly quantized vortex in the P phase is computed numerically.
We assume a straight vortex along the $z$ axis and solve the time-independent GP equations, $0=\frac{\delta G}{\delta\Psi_m}$, in two dimensions.
Figure~\ref{Fig-Cross} shows the typical structures of a vortex core.
Here, the core structures in the vicinity of the vortex axis at $(x,y)=(0,0)$ are demonstrated, and
 we solved the GP equation with a cylindrical potential whose radius $R$ is much larger than the relevant lengths, $R\approx 243 \xi_n$.
The method of numerical computation is the same as that in Ref.~\cite{takeuchi2020quantum}.

A vortex with the lowest energy is classified into three types depending on the two variables $\frac{q}{\mu}$ and $\frac{c_s}{c_n}$.
The simplest type of vortex is a normal(N)-core vortex,
 whose core is vacant by forming ``the normal (N) state'' with $\Psi_m=0$ at the vortex axis.
An N-core vortex has the lowest energy when $\frac{q}{\mu}$ exceeds a critical value $\frac{q_{\rm C}}{\mu}$, which is a function of $\frac{c_s}{c_n}$.
For $q<q_{\rm C}$ and $c_s>0$,
 the vortex axis is occupied by the AF state with a nonaxisymmetric density profile forming an elliptic velocity field, as was found in Ref.~\cite{takeuchi2020quantum}.
 We call this vortex an elliptic AF-core vortex [Fig.~\ref{Fig-Cross}(a)], or more simply, an AF-core vortex in this paper.
The (axisymmetric) F-core vortex has the lowest energy for $-2\frac{c_s}{c_n}<\frac{q}{\mu}<\frac{q_{\rm C}}{\mu}$ and $c_s<0$.
The vortex axis of an F-core vortex is magnetized with $s_z=\pm |\Psi_{\pm 1}|^2$ and $|\Psi_{\mp 1}|^2=0$, and its density profile is axisymmetric [Fig.~\ref{Fig-Cross}(b)].

The transition between AF- and F-core vortices is discontinuous.
An AF(F)-core vortex can be realized as a metastable vortex in certain parameter regimes in the numerical simulations for $c_s<0$ ($c_s>0$),
in which the difference in the vortex energy between the lowest-energy vortex and the metastable one is not so large.
Metastable F-core vortices were observed as vortices with finite magnetization at the core in the later stage of quenched phase transition dynamics in the experiment \cite{kangPhysRevLett.122.095301}.
It is implied that the probability of the appearance of metastable vortices in quench dynamics will increase for larger $\frac{q}{\mu}$ and smaller $\frac{c_s}{c_n}$
because of the fact that the energy difference increases with decreasing $\frac{q}{\mu}$ or increasing $\frac{c_s}{c_n}$.
The energy difference is attributed mainly to the spin interaction energy;
an F-core vortex is largely magnetized compared with an AF-core vortex,
which makes the vortex energy of an F-core vortex smaller (larger) than that of an AF-vortex for negative (positive) $c_s$.

In contrast, the transition from an N-core vortex to an AF-core vortex or an F-core one is continuous.
The transition is realized by the appearance of the $m=\pm 1$ component in the vortex core,
 which is preferred energetically for smaller $\frac{q}{\mu}$.
This transition could be understood as the occurrence of another ordered state at the core of the topological defect
by analogy with the phase transitions of solitons in the P phase \cite{liu2020phase} and nematic-spin vortices in the AF phase \cite{underwood2020properties}.
In such a transition, where the spin interaction is negligible with $|c_s| \ll c_n$ or ${\bm s}=0$,
the relevant length scales are the (density) healing length $\xi_n$ and the Zeeman length $\xi_q$.
Especially speaking, for $q\ll \mu$, the spatial variation of the density is negligible
and thus the size of the topological defects is determined by the balance between the quadratic Zeeman energy and the gradient (kinetic) energy associated with the pseudo-director field.
 Actually, we found that
the $m=\pm 1$ component occupies the vortex core, such that the total density $n$ is nearly homogeneous for small $\frac{q}{\mu}$ in the cross-section profiles (not shown).

The vortex becomes an N-core vortex when the core size $\xi_q$ is comparable to $\xi_n$ for $\frac{|c_s|}{c_n}\ll 1$,
 giving the critical point $\frac{q_{\rm C}}{\mu}\approx 0.25$ according to Ref.~\cite{takeuchi2020quantum}.
The dependence of the critical point on the spin interaction is qualitatively understood as follows.
In general, the energy of a state with magnetization increases with increasing $c_s$.
Therefore, a vortex state with magnetization at the core is more preferred for smaller $c_s$ energetically.
Thus, the transition point $\frac{q_{\rm C}}{\mu}$ is a decreasing function of $\frac{c_s}{c_n}$ as is shown in the phase diagram.

\section{Axisymmetric vortex}\label{sec-AV}
The properties of an axisymmetric vortex are investigated analytically.
A singly quantized vortex in the P phase is axisymmetric for $-\frac{q}{2\mu} <\frac{c_s}{c_n}\leq 0$ or $q\geq q_{\rm C}$, including the critical point of $c_s=0$.
We derive the vortex winding rule from the general formalism of an axisymmetric vortex in spin-1 BECs,
 which is also important to understand the axisymmetry breaking of elliptic vortices.
The hydrostatic approximation is introduced to qualitatively describe the cross-section profiles of the vortex core.

\subsection{Vortex winding rule}

First, we discuss the stationary solutions of an axisymmetric vortex by starting from the following ansatz
\begin{eqnarray}
\Psi_m=f_m(\rho)e^{i(L_m\varphi+\vartheta_m)},
   \label{eq:Avortex}
\end{eqnarray}
with the radius $\rho=\sqrt{x^2+y^2}$ and the azimuthal angle $\varphi$ in cylindrical coordinates.
Here, the winding number $L_m$ is an integer, $\vartheta_m$ is a real constant, and $f_m(\rho)$ is a real function.
By substituting Eq.~(\ref{eq:Avortex}) into the time-independent GP equations $0=\frac{\delta G}{\delta \Psi_m}$, we obtained the following coupled radial equations:
\begin{eqnarray}
&& 0=h_{0} f_{0} +c_{s}f_{0}\left( f^{2}_{+1} +f^{2}_{-1} +2f_{+1} f_{-1} e^{i\delta \Theta }\right),
\label{eq:f_0r}\\
&& 0= h_{\pm1} f_{\pm1} +c_{s} f^{2}_{0}\left( f_{\pm 1} +f_{\mp 1} e^{-i\delta \Theta }\right),
   \label{eq:f_pmr}
\end{eqnarray}
with
\begin{eqnarray}
 h_{m} &=&  \frac{\hbar^{2}}{2M}\left(-\frac{{\rm d}^{2}}{{\rm d}\rho^{2}} -\rho^{-1}\frac{{\rm d}}{{\rm d}\rho}+\frac{ L^{2}_{m}}{\rho^{2}}\right).
\nonumber \\
 &&-\mu+m^2q-mp+c_nn+mc_ss_z.
 \label{eq:hm}
\end{eqnarray}

According to Eqs.~(\ref{eq:f_0r}) and (\ref{eq:f_pmr}),
$\delta\Theta$ must be an integer multiple of $\pi$ for $f_{+1}f_{-1}f_0\neq 0$ with $c_s\neq 0$.
Hence, the winding numbers satisfy the vortex winding rule as follows:
\begin{eqnarray}
 L_m=L+mN.
 \label{eq:winding}
\end{eqnarray}
with integers $L(=L_0)$ and $N$, which are related to the mass- and spin-currents around the vortex.
Here, we set $\vartheta_{+1}+\vartheta_{-1}-2\vartheta_{0}=0$ in general,
because $\delta \vartheta=\pi$ is reproduced by changing the sign of $f_m$, for example, $f_{+1}\to -f_{+1}$ or $f_{-1}\to -f_{-1}$ with $\arg\Psi_0=0$.
Assuming these conditions, Eqs.~(\ref{eq:f_0r}) and (\ref{eq:f_pmr}) are reduced to
\begin{eqnarray}
&& 0=h_{0}f_{0} +c_{s}f_{0}( f_{+1} +f_{-1})^2,
\label{eq:f0}\\
&& 0=h_{\pm1} f_{\pm1} +c_{s} f^{2}_{0}( f_{+ 1} + f_{- 1}).
  \label{eq:fpm}
\end{eqnarray}
The amplitudes $f_m$ in the axisymmetric vortex states obey these equations.

\subsection{Local phase transition in a coreless vortex}

Here, we introduce a qualitative analysis to describe the core structure of a vortex in the P phase.
In the core of a singly quantized vortex in scalar BECs,
the order parameter vanishes by forming a normal (disordered) state at the vortex axis ($\rho=0$).
One might expect the same thing to happen for a vortex in the P phase, in which the order parameter is a scalar complex field $\Psi_0$, as is the case for scalar BECs.
However, the core may be occupied by other spin components to reduce the energy, where the order parameter $\Psi_{+1}$ and/or $\Psi_{-1}$ have a finite amplitude at the vortex axis.

This is typical in multicomponent superfluids.
Such a vortex is sometimes called a coreless vortex.
Note that we should not confuse coreless vortices with continuous vortices in a narrow sense.
According to the literature of superfluid $^3$He \cite{vollhardt2013superfluid},
the total amplitude of the multicomponent order parameter is homogeneous in a continuous vortex;
 that is, the ordered state in the bulk is the same as that in the core except for the ``orientation'' of the order parameter.
 The most famous example of a continuous vortex is the Mermin--Ho vortex,
  the vortex-core structure of which is expressed as a spatial variation of the orientation of the vector order parameter, called the Mermin--Ho texture \cite{vollhardt2013superfluid}.
 A vortex is categorized as a singular vortex when the bulk ordered state is not realized in the vortex core.
All vortices in the P phase are singular vortices in the sense that ordered states other than the bulk P state are realized in the vortex core.

The concept of the coreless vortex is more general than that of the continuous vortex.
A coreless vortex is a quantized vortex whose vortex core is occupied by other ordered states.
Therefore, the AF- and F-core vortices are coreless vortices.
A change in the internal state in the vortex core can be understood as an occurrence of local phase transition, because the ordered state in the vicinity of the vortex axis is different from that in the bulk.
We evaluate such phase transition for an axisymmetric vortex in an approximate model below.

\subsection{Hydrostatic approximation}

Although the N-core vortex is not categorized as a coreless vortex,
we first consider the local phase transition in a conventional vortex in scalar BECs as the simplest case.
The vortex state is described by the following equation, as obtained from Eq.~(\ref{eq:f0}) with $f_{+1}=f_{-1}=0$,
\begin{eqnarray}
 0=\left[ \frac{\hbar ^{2}}{2M}\left(-\frac{{\rm d}^{2}}{{\rm d}\rho^{2}} -\rho^{-1}\frac{{\rm d}}{{\rm d}\rho}\right)
 -\bar{\mu}+c_nf_0^2 \right]f_{0}
\label{eq:Nvor}
\end{eqnarray}
with
\begin{eqnarray}
  \bar{\mu}=\mu-\frac{\hbar^{2}L^2}{2M\rho^{2}}.
  \label{eq:mub}
\end{eqnarray}
This quantity is called the hydrostatic chemical potential, which is named after the (hydro)static pressure in the context of the Bernoulli theorem in quantum hydrodynamics \cite{tsubota2013quantum}.
Accordingly, we call the first and second terms on the right hand side of Eq.~(\ref{eq:mub}) the total and (hydro)dynamic chemical potentials, respectively.

In the hydrostatic approximation, by neglecting the radial gradient of $f_0$,
the local equilibrium state is determined by the hydrostatic chemical potential $\bar{\mu}$.
Although we have $f_0=0$ for nonpositive $\bar{\mu}$,
 the order parameter has a finite amplitude with $f_0 \geq 0$ in the approximation for $\bar{\mu} \geq 0$,
\begin{eqnarray}
   f_0(\rho)=\sqrt{\frac{\bar{\mu}(\rho)}{c_n}}~~\left(r\geq \frac{|L|\xi_n}{\sqrt{2}}\right).
   \label{eq:f0TF}
\end{eqnarray}
The result suggests that a normal state is realized in the vortex core by forming an N-core vortex.

\begin{figure}
 \begin{center}
 \includegraphics[width=1.0 \linewidth, keepaspectratio]{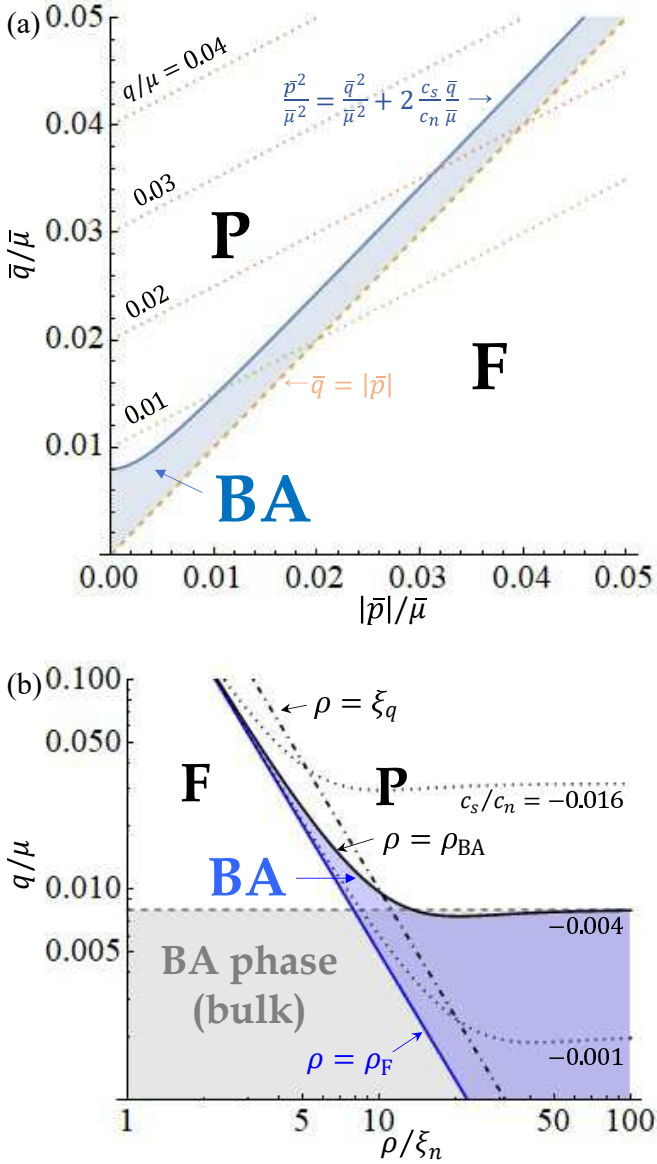}
 \end{center}
 \vspace{-5mm}
 \caption{
 (a) Bulk phase diagram of spin-1 BECs for $\frac{c_s}{c_n}=-0.004$.
 The chemical potential $\mu$ and the Zeeman coefficients $q$ and $p$ are replaced by their hydrostatic counterparts $\bar{\mu}$, $\bar{q}$, and $\bar{p}$, respectively.
 The dotted lines correspond to the parametric equations of the position $(\bar{p}/\bar{\mu},\bar{q}/\bar{\mu})$ in the phase diagram for $\frac{q}{\mu}=0.01$, $0.02$, $0.03$, and $0.04$ with $p=0$,
  where the parameter is $\frac{\rho}{\xi_n}$.
 (b) Radial profile of the local phase transition in an F-core vortex for $c_s/c_n=-0.004$ and $p=0$, corresponding to the mapping $(\bar{p}/\bar{\mu},\bar{q}/\bar{\mu})\to(q/\mu, \rho/\xi_n)$ from (a) in the hydrostatic approximation.
 Solid curves represent $\rho=\rho_{\rm F}$ (blue) and $\rho=\rho_{\rm BA}$ (black), respectively.
 The P phase is unstable and the BA phase is energetically preferred in the bulk when $\frac{q}{\mu}$ is smaller than $-2\frac{c_s}{c_n}$ (gray dashed line).
 The gray, shaded area corresponds to the parameter region, where the BA phase is realized in the bulk.
The dash-dotted line represents $\rho=\xi_q$.
 The F phase is realized in the vicinity of the vortex axis (F for $\rho<\rho_{\rm F}$).
 The F core is surrounded by the BA skin ($\rho_{\rm F} < r < \rho_{\rm BA}$), in which the local BA phase occurs.
 The P phase is the lowest energy state in the bulk and is realized for $\rho>\rho_{\rm BA}$.
 The F-core vortex is never realized for a large $\frac{q}{\mu}$, where $\rho_{\rm BA}$ and $\rho_{\rm F}$ are smaller than $\rho_{\rm N}$, and an N-core vortex is realized.
 The dotted curves indicate $\rho=\rho_{\rm BA}$ for $\frac{c_s}{c_n}=-0.016$ (top) and $-0.001$ (bottom).
 The BA skin becomes thinner for smaller negative values of $\frac{c_s}{c_n}$.
 }
 \label{Fig-PDbulk}
 \end{figure}

Next, we take into account the contribution of the $m=\pm 1$ components more generally.
To describe the phase transition of the vortex core systematically,
we define the hydrostatic Zeeman coefficients as
\begin{eqnarray}
&&
\bar{q} =q+\frac{\hbar ^{2} N^{2}}{2M\rho^{2}},
\label{eq:qb}\\
&&
\bar{p} =p-\frac{\hbar ^{2} LN}{M\rho^{2}}.
\label{eq:pb}
\end{eqnarray}
By substituting Eqs.(\ref{eq:mub}--\ref{eq:pb}) into Eqs.(\ref{eq:f0}) and (\ref{eq:fpm}),
 and neglecting the gradient terms of $f_m$,
the hydrostatic approximation yields
\begin{eqnarray}
&& 0=\bar{h}_{0}f_{0} +c_{s}f_{0}( f_{+1} +f_{-1})^2,
\label{eq:f0a}\\
&& 0=\bar{h}_{\pm1} f_{\pm1} +c_{s} f^{2}_{0}( f_{+1} +f_{-1}),
  \label{eq:fpa}
\end{eqnarray}
with
\begin{eqnarray}
 \bar{h}_{m} =  -\bar{\mu}+m^2\bar{q}-m\bar{p}+c_nn+mc_ss_z.
 \label{eq:hmshift}
\end{eqnarray}
These equations have the same form as Eqs.~(\ref{eq:f0}) and (\ref{eq:fpm}) after the replacement $h_m\to\bar{h}_m$.
In this approximation, the local value of $f_m(\rho)$ is obtained from the bulk solution with $(\mu,q,p)$ replaced by $(\bar{\mu},\bar{q},\bar{p})$.
Therefore, we can describe the ordered state in the vortex core according to the phase diagram of spin-1 BECs in the bulk, as shown in Fig.~\ref{Fig-PDbulk}(a) (see also Fig.~3(c) in Ref.~\cite{kawaguchi2012spinor}).

Now, the vortex solutions are characterized by three dimensionless variables, $\frac{\bar{q}}{\bar{\mu}}$, $\frac{c_s}{c_n}$, and $\frac{\bar{p}}{\bar{\mu}}$.
The ordered state around the vortex core can change depending on the distance $\rho$ from the vortex axis in the hydrostatic approximation,
because the three variables,$\frac{\bar{q}}{\bar{\mu}}$, $\frac{c_s}{c_n}$, and $\frac{\bar{p}}{\bar{\mu}}$, are expressed as functions of $\frac{\rho}{\xi_n}$.
Here, we restrict ourselves to the case of $c_s \leq 0$
because a nonaxisymmetric vortex is energetically preferred for antiferromagnetic interaction $c_s>0$
except for the case of $q>q_{\rm C}$, for which an N-core vortex is realized as the lowest-energy vortex.

\subsection{Ferromagnetic interaction ($c_s<0$)}

We consider an axisymmetric vortex of $L=1$ for $p=0$.
The lowest energy state is investigated for $N=0, \pm 1$.
For ferromagnetic interaction $0>\frac{c_s}{c_n}>-\frac{q}{2\mu}$ in the P phase,
the local phase transition from the P phase to the BA phase occurs with
\begin{eqnarray}
 \left(\frac{\bar{p}}{\bar{\mu}}\right)^2=\left(\frac{\bar{q}}{\bar{\mu}}\right)^2+2\frac{c_s}{c_n}\frac{\bar{q}}{\bar{\mu}}~~~({\rm P-BA~transition}).
 \label{eq:TPBA}
\end{eqnarray}
This condition is satisfied at $\rho=\rho_{\rm BA}$.

For $N=\pm 1$,
  the radius $\rho_{\rm BA}$ is given by
\begin{eqnarray}
 \frac{\rho_{\rm BA}^2}{\xi_n^2}
 =\frac{\sqrt{8\tilde{c}_s\tilde{q}+4\tilde{q}^2+\tilde{c}_s^2\left(1+\tilde{q}\right)^2}+\tilde{c}_s\tilde{q}-\tilde{c}_s-\tilde{q}}{2\tilde{q}(2\tilde{c}_s+\tilde{q})}
 \label{eq:rPBA}
\end{eqnarray}
with $\tilde{c}_s=\frac{c_s}{c_n}$ and $\tilde{q}=\frac{q}{\mu}$.
In addition to the P--BA transition at $\rho=\rho_{\rm BA}$,
 the phase transition from the BA phase to the F phase occurs at $\rho=\rho_{\rm F}<\rho_{\rm BA}$.
The transition radius is given by the relation
\begin{eqnarray}
\bar{q}=|\bar{p}|~~~({\rm BA-F~transition}),
 \label{eq:TPF}
\end{eqnarray}
which yields
\begin{eqnarray}
 \frac{\rho_{\rm F}^2}{\xi_n^2}=\frac{\mu}{2q}.
 \label{eq:rBAF}
\end{eqnarray}

For the case of $(L,N)=(1,0)$, we have solutions of neither $\rho_{\rm N}<\rho_{\rm BA}$ nor $\rho_{\rm N}<\rho_{\rm F}$.
This means that an N-core vortex is always realized in this case.
This is reasonable, because all components vanish at the vortex axis owing to the centrifugal term with $L_{0,\pm 1}\neq 0$.
In fact, we have never obtained such vortex states as the lowest-energy solution for $q<q_{\rm C}$ in the numerical simulations.
Hence, we consider the case of $(L,N)=(1,\pm 1)$ below.

Figure~\ref{Fig-PDbulk}(b) shows the radial distribution of the local ordered states in the vortex core for $\frac{c_s}{c_n}=-0.004$ with $p=0$, as obtained by the hydrostatic approximation.
A similar value of $\frac{c_s}{c_n}$ is realized for spin-1 BECs of $^{87}$Rb.
The results of Eqs.~(\ref{eq:rPBA}) and (\ref{eq:rBAF}) are displayed as black and blue solid curves, respectively.
The approximation suggests the existence of the local BA phase as the ``BA skin'' surrounding the local F phase in the vortex core
[see also the schematic of an F-core vortex (Fc) in Fig.~\ref{Fig-diagram}].
The vortex-core structure is understood more easily from the bulk phase diagram.
The four dotted lines in Fig.~\ref{Fig-PDbulk}(a) correspond to the parametric plots of $\left(\frac{\bar{q}(\rho)}{\bar{\mu}(\rho)}, \frac{|\bar{p}(\rho)|}{\bar{\mu}(\rho)} \right)$ from $\rho=0$ to $\rho=\infty$ for $\frac{q}{\mu}=0.01,~0.02,~0.03,~0.04$ with $p=0$.
The bulk P phase is realized for $\frac{|\bar{p}|}{\bar{\mu}}\to 0$ ($\rho\to \infty$).
The BA skin appears by increasing $\frac{|\bar{p}|}{\bar{\mu}}$ (decreasing $\rho$), and the local state is finally in the local F phase for $|\bar{p}|>\bar{q}$.

The presence of the BA skin between the local F and P phases implies that the spin density ${\bm s}$ lies in the $xy$ plane for large $\frac{\rho}{\xi_n}=\sqrt{\frac{\mu}{|\bar{p}|}}$ with $p=0$.
This is because, for $\rho \gg \xi_n$, $s_z$ is proportional to $\bar{p}$ and becomes negligibly small compared with $s_\bot$ in the approximation.
Moreover, the existence of the F core is topologically protected by the vortex winding rule;
according to Eq.~(\ref{eq:winding}) with $L=1$ and $N=\mp 1$,
the $m=\pm 1$ component can take a finite amplitude at the vortex axis with $L_{\pm 1}=0$
whereas $f_{\mp 1}$ and $f_{0}$ must be zero because of the centrifugal potential $\propto \frac{L_m^2}{\rho^{2}}$ in Eq.~(\ref{eq:hm}).
Then, we have the magnetization with $s_z=\pm f_{\pm 1}^2=\pm n$ and $s_\bot =0$ at $\rho=0$.

This suggests that the spin texture in the vortex core can be similar to that of the Mermin--Ho vortex \cite{vollhardt2013superfluid};
the orientation of the spin in the texture sweeps a semisphere in the spin space.
In the Mermin--Ho vortex,
the amplitude of the order parameter is homogeneous, and thus the texture is represented by a unit vector field throughout the system.
In contrast, however, the texture of an F-core vortex is localized and disappears for $\frac{\rho}{\xi_n} \to \infty$  [see Fig.~\ref{Fig-Cross}(b) for $N=-1$].
This is because the order parameter in the core is different from that in the bulk, where the spin density disappears far from the vortex core.

\begin{figure}
\begin{center}
\includegraphics[width=1.0 \linewidth, keepaspectratio]{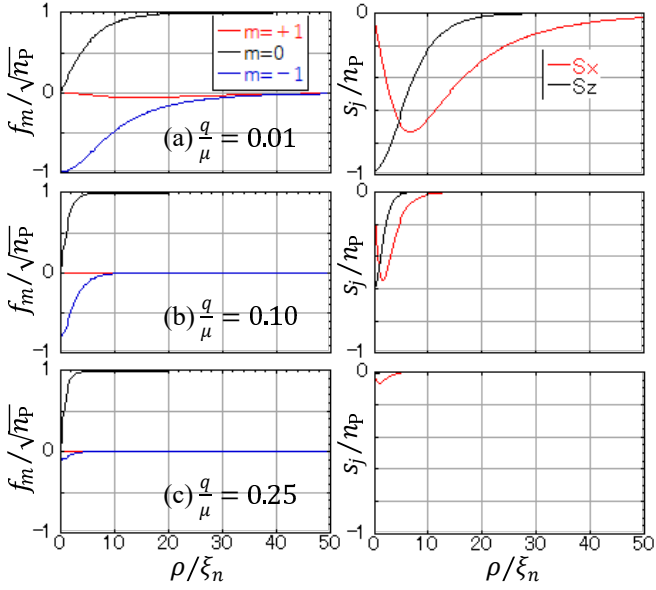}
\end{center}
\vspace{-5mm}
\caption{
Numerical results of the radial profiles of $f_m$ (left) and $s_j~(j=x,y,z)$ (right) in an F-core vortex of $L=N=1$ for $\frac{c_s}{c_n}=-0.004$ with (a) $\frac{q}{\mu}=0.01$, (b) $\frac{q}{\mu}=0.10$, and (c) $\frac{q}{\mu}=0.25$.
The $m=-1$ component has a finite amplitude at the vortex axis ($\rho=0$), forming an F-core vortex for $q<q_{\rm C}=0.2522$, whereas an N-core vortex with $\Psi_m(\rho=0)=0$ is formed for $q>q_{\rm C}$.
}
\label{Fig-FcoreV}
\end{figure}

To investigate the core structure more precisely,
we obtained the radial profiles of $f_m$ and ${\bm s}$ by numerically solving Eqs.~(\ref{eq:f_0r}) and (\ref{eq:f_pmr}) for $L=N=1$,
\footnote{
The steepest descent method is employed to obtain the vortex solution numerically.
The radial coordinate $\rho$ is discretized as $\rho \to \rho_i=i\Delta \rho~(i=0,1,2,\cdots)$ with a grid size $\Delta \rho=0.25\xi_n$.
The radial derivatives of $f_m$ are computed with the finite difference approximation; for example, $\frac{{\rm d}}{{\rm d}\rho}f_m$ and $\frac{{\rm d}^2}{{\rm d}\rho^2}f_m$ are computed by the central difference of the first and second orders, respectively.
}.
Figure~\ref{Fig-FcoreV} shows the numerical results for several values of $\frac{q}{\mu}$ with $\frac{c_s}{c_n}=-0.004$.
For small $\frac{q}{\mu}$,
the numerical result is consistent with the prediction by the hydrostatic approximation.
In this case, the vortex core is approximately divided into two regions associated with the spin distribution.
A region around the vortex axis ($\rho=0$) is in the F state, where $|s_z|$ is larger than $s_\bot$, corresponding to the F core.
The other is in the BA state with $s_\bot >|s_z|$, which forms the BA skin.
The P state with $s_\bot \ll n_{\rm P}$ and $|f_0|\gg |f_{\pm 1}|$ is present outside the BA skin.
The boundaries between these different states are not clearly visible in the numerical plots, owing to the penetration effect of the macroscopic wave function.
This effect is described by the gradient term of $f_m$, which is neglected in the hydrostatic approximation.

To understand the penetration effect beyond the hydrostatic approximation, we consider the asymptotic behavior of the spin density.
This effect is described by the penetration of $f_{\pm 1}$ in the deep P-state region ($\rho\gg \rho_{\rm BA}$)
 as $g_{\pm}\equiv \pm \sqrt{\frac{c_n}{2\mu}}(f_{+1}\pm f_{-1})\sim -\sqrt{\frac{\xi_n}{\rho}}e^{-\frac{\rho}{\xi_{\pm}}}$ according to Eq.~(\ref{eq:f_pmr}),
  where we used $\xi_+= \xi_q/\sqrt{2+4\frac{c_s\mu}{c_nq}}$ and $\xi_-=\xi_q/\sqrt{2}$.
As a result, the penetration depths of $s_\bot\approx -2n_{\rm P}g_+ \propto e^{-\frac{\rho}{\xi_\bot}}$ and $s_z=-2n_{\rm P} g_+g_-\propto e^{-\frac{\rho}{\xi_z}}$ are given by
\begin{eqnarray}
&&\xi_\bot=\xi_+=\frac{1}{\sqrt{2}}\xi_q,
\label{eq:xi_bot}\\
&&\xi_z=\left( \frac{1}{\xi_+}+\frac{1}{\xi_-} \right)^{-1}=\left(1+\sqrt{1+2\frac{c_s/c_n}{q/\mu}}\right)^{-1}\xi_\bot.
\end{eqnarray}
The presence of the BA skin is justified by the fact that $\xi_\bot$ is always larger than $\xi_z$.
This result also explains the behavior of the core structure near the critical point for $q \sim q_{\rm C}$,
where the hydrostatic approximation breaks down with $\rho_{\rm BA}\sim \xi_n$;
the amplitude of $s_z$ decreases with increasing $\frac{q}{\mu}$ more rapidly than $s_\bot$, and
 $s_z$ becomes negligibly small in the critical regime, as shown in Fig.~\ref{Fig-FcoreV}(c).

The core size of an F-core vortex is characterized by Zeeman length $\xi_q$ because the penetration depth of Eq.~(\ref{eq:xi_bot}) is understood as the radius of the BA skin according to the asymptotic behavior. This conclusion is also supported by the hydrodynamic approximation. To compare $\xi_\bot$ with the hydrostatic approximation, we plot $\rho=\xi_q$ as a dash-dotted curve in Fig.~\ref{Fig-PDbulk}(b) as reference. We can see that the hydrostatic approximation ($\rho=\rho_{\rm BA}$) is supported by the above conclusion quantitatively except near the transition point of the bulk BA phase.

\subsection{Critical point ($c_s=0$)}\label{subsec-CP}

It is instructive to consider the critical point $c_s=0$,
across which the lowest-energy vortex changes from the F-core vortex to the AF-core vortex.
The winding rule is not applicable at the critical point,
 because the terms of the exponent $e^{\pm i\delta\Theta}$ in Eqs.~(\ref{eq:f_0r}) and (\ref{eq:f_pmr}) become zero with $c_s=0$.
Then, we may assume an axisymmetric vortex with, for example, $(L_{+1},L_0,L_{-1})=(0,1,0)$.
In this vortex state, both $f_{+1}$ and $f_{-1}$ may be finite at the vortex axis.
Because the energy density ${\cal U}$ depends on $\Psi_{\pm 1}$ only through the term $n_\bot=f_{+1}^2+f_{-1}^2$ with $c_s=p=0$,
the energy of the axisymmetric vortex is invariant under the rotation of the unit vector field $\frac{1}{\sqrt{n_\bot}(\rho)}[f_{+1}(\rho),f_{-1}(\rho)]^{\rm T}$,
which does not change the radial profile of $n_\bot(\rho)$.

An operation of the vector rotation realizes an AF-core vortex with $f_{\pm1}^2=\frac{n_\bot}{2}$,
where ${\bm s}({\bm r}=0)=0$ at the vortex axis forming the local AF phase.
 Similarly, the local F phase occurs for $f_{\pm 1}^2=n_\bot$ and $f_{\mp 1}(\rho)=0$, forming an F-core vortex.
Therefore, the AF-core vortex has the same energy as the F-core vortex at the critical point of $c_s=0$.
The local BA phase never occurs at the vortex axis, because $s_\bot$ vanishes there with $\Psi_0=0$.

The above argument implies that the $m=\pm 1$ components vanish in the F-core vortex with $(L,N)=(1,\pm 1)$ when the parameter $\frac{c_s}{c_n} (<0)$ approaches the critical point, $\frac{c_s}{c_n}\to 0$.
 This conjecture is consistent with the fact that the amplitude of $m=+1$ is very small in Fig.~\ref{Fig-FcoreV} with a small negative $c_s$.
 It was also found numerically that the amplitude increases with increasing $|c_s|$ (not shown).
 The vector rotation transforms the vortex state from the F-core vortex into the AF-core vortex with axisymmetry without an energy change at the critical point.
 Then, we can say that the elliptic vortex on the positive-$c_s$ side recovers axisymmetry as $\frac{c_s}{c_n}$ is close to zero.
 The core size of the AF-core and F-core vortices at the critical point is characterized by $\xi_q$,
because  we have $\rho_{\rm BA}=\rho_{\rm F}\sim \xi_q$ with $c_s=0$ in the hydrostatic approximation.

As was mentioned before,
the vortex-core transition at the critical point is discontinuous and
one of the two types of vortices is the lowest-energy vortex or the metastable vortex for finite $c_s$.
The former claim is justified by the fact that we require an operation of the vector rotation by $\frac{\pi}{4}$ for the transformation from an AF-core vortex to an F-core vortex at the critical point while the operation must be an infinitesimal rotation for continuous transition.
The latter is understood by considering the contribution of the spin interaction as follows.
The transverse magnetization is broadly distributed in the F-core vortex, whereas it is localized at the BA edges of the AF-core vortex.
In addition, the longitudinal spin density is present around the vortex axis in the F-core vortex.
The energy contribution from the spin interaction is relatively large in the F-core vortex,
 and hence the AF-core vortex is the lowest-energy vortex and the metastable vortex for $c_s>0$ and $c_s<0$, respectively.

\section{Elliptic vortex}\label{sec-EV}

An elliptic vortex is the lowest-energy vortex in the P phase for $\frac{c_s}{c_n}>0$ and small $\frac{q}{\mu}$.
This vortex is no longer axisymmetric by forming an elliptic velocity field with a planar singularity in vorticity.
In Ref.~\cite{takeuchi2020quantum}, an elliptic vortex is evaluated by applying the Joukowski transformation to an axisymmetric vortex.
Here, we demonstrate a qualitative description of the distribution of the ordered state in the core of an elliptic vortex from a viewpoint different from the quantitative analysis in Ref.~\cite{takeuchi2020quantum}.
The hydrostatic approximation demonstrated above is applied after some modification according to the transformation.

To describe the core structure of an elliptic vortex qualitatively,
we assume an axisymmetric state of $L_0=1$ and $L_{+1}=L_{-1}=0$ in Eq.~(\ref{eq:Avortex}) with $\vartheta_0=0$.
This consideration gives us an intuitive approach for understanding the elliptic structure,
 although the state is not realistic by breaking the vortex winding rule.
Additionally, we consider that the local AF phase is realized at the axis of this vortex with $\Psi_0(\rho=0)=0$,
 and then set $\vartheta_{+1}+\vartheta_{-1}=\pi$ to realize $s_\bot=0$ there, yielding
\begin{eqnarray}
\delta\Theta=2\varphi+\pi.
\end{eqnarray}
This state is not axisymmetric in the sense that the transverse spin density of Eq.~(\ref{eq:sbot}) depends on the azimuthal angle $\varphi$.
The magnetization is minimized and maximized at $\varphi=0,\pi$, and $\varphi=\pm \frac{\pi}{2}$, respectively.
This vortex state can be regarded as the consequence of the inverse of the Joukowski transformation from an elliptic vortex.

To highlight the essential points,
we consider the profiles of the order parameters along the $x$ axis and $y$ axis in the vortex,
which are similar to the AF-core and BA-core soliton of Ref.~\cite{liu2020phase}, respectively.
In the profile along the $x$ axis,
the wave function of the $m=0$ component is real, $\Psi_0(x,y=0)={\rm sgn}(x)f_0(\rho=|x|)$, and the real function changes its sign at $x=0$, forming a structure such as a dark soliton.
The core of the soliton is occupied by the $m=\pm 1$ components with $\Psi_{+1}(y=0)=f_{+1}$ and $\Psi_{-1}(y=0)=f_{-1}=-f_{+1}$.
This structure is the same as the core structure of the AF-core soliton found in Ref.~\cite{liu2020phase}.
Accordingly, the local AF phase is realized at the vortex axis ($x=y=0$).

The wave function is written as $\Psi_0(x=0,y)={\rm sgn}(y)f_0(\rho=|y|)i$ along the $y$ axis, and the real function $i\Psi_0$ forms a dark soliton.
The transverse spin density is nonzero near the center of the soliton core at $y=0$ with $s_y={\rm sgn}(y)\sqrt{2}f_0(f_{+1}-f_{-1})$.
This profile is physically identical to that of the BA-core soliton \cite{liu2020phase} when ${\bm s}$ is rotated about the $z$ axis by $\pi/2$.
Correspondingly, the local BA phase occurs near the vortex axis along the $y$ axis.
This transverse magnetization corresponds to the two spin spots as the BA edges of the elliptic vortex.

The local AF phase is energetically preferred over the local BA phase for antiferromagnetic interactions.
In this sense, the BA edges are regarded as the {\it by-products} of the AF core,
 which could occur owing to the winding of the phase $\arg \Psi_0$.
 This energetic argument can be restated in terms of the soliton energy as follows:
 the AF-core soliton is energetically preferred over the BA-core soliton for $c_s>0$.
This is a qualitative explanation of why the AF-core region is elongated along the $y$ axis in the elliptic vortex [see Fig.~\ref{Fig-Cross}(a)] by forming an AF-core soliton of a finite length while the length of the BA-core soliton becomes ``zero'' to avoid an unnecessary energy cost.

The length of the AF-core soliton is determined by a balance between the soliton energy and the hydrodynamic potential induced by the elliptic velocity field \cite{takeuchi2020quantum}.
Therefore, the soliton length between the two spin spots depends on the tension coefficient of the soliton.
The tension coefficient is an increasing function of $\frac{q}{\mu}$ and takes its highest value for $q>q_{\rm C}$,
 where the soliton becomes a dark soliton with $\Psi_{\pm 1}=0$.
Then, the soliton length is zero and forms an N-core vortex.

\begin{figure*}
\begin{center}
\includegraphics[width=1.0 \linewidth, keepaspectratio]{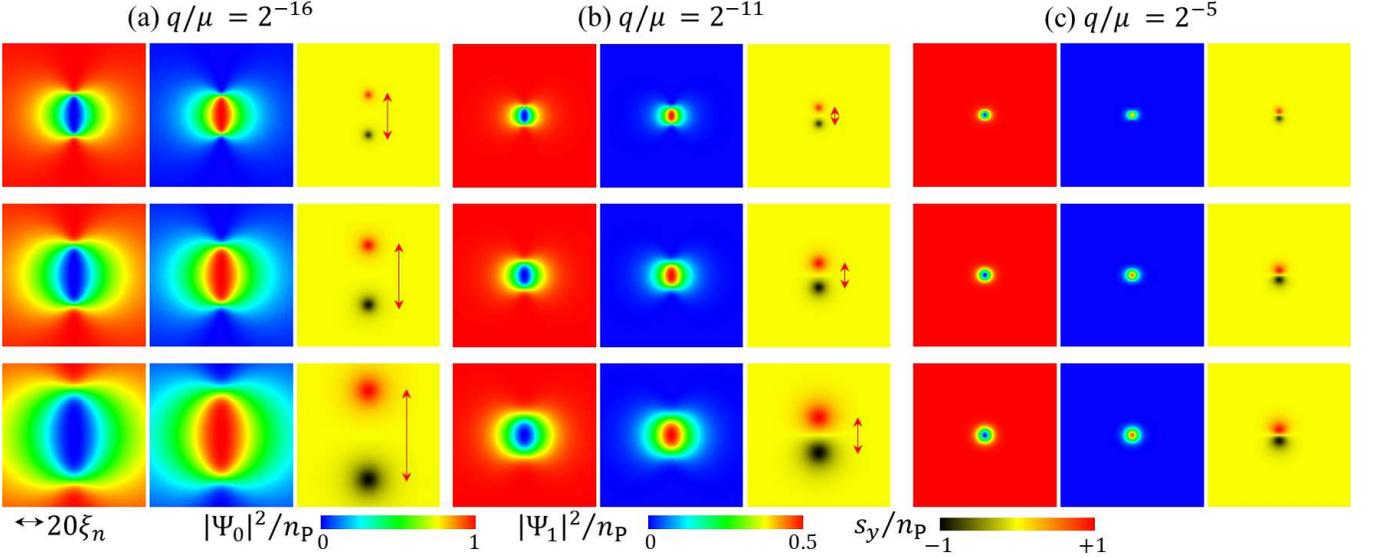}
\end{center}
\vspace{-5mm}
\caption{
Cross-section plots of $|\Psi_0|^2$ (left), $|\Psi_{1}|^2$ (center), and $s_y$ (right) of the AF-core vortices for (a) $q/\mu= 2^{-16}$, (b) $q/\mu= 2^{-11}$, and (c) $q/\mu= 2^{-5}$.
The top, middle, and bottom panels show the plots of $c_s/c_n=0.128$, $0.032$, and $0.008$, respectively.
The red arrow represents the distance between the two spin spots estimated by the phenomenological model of Eq.~(\ref{eq:dxin}).
Each plot is displayed on the same scale as is indicated by a double-headed arrow in the lower left.
}
\label{Fig-Evortex}
\end{figure*}

The size (radius) of a spin spot is uniquely characterized by the spin healing length for small $\frac{q}{\mu}$, such that $\xi_q \gg \xi_s$, where the density $n$ is approximately homogeneous.
Here, the healing length is estimated as
\begin{eqnarray}
\xi_s=\frac{\hbar}{\sqrt{M|c_s|n_{\rm P}}}.
\end{eqnarray}
Figure~\ref{Fig-Evortex} shows the cross-section plots of AF-core vortices for several values of the parameters.
The spin healing lengths for $\frac{c_s}{c_n}=0.128$ (top), $0.032$ (middle), and $0.008$ (bottom) are given by $\frac{\xi_s}{\xi_n}\approx 2.8$, $5.6$, and $11.2$, respectively.
We can see that the size of the spin spots consistently decreases with $\xi_s$.

The width of the elliptic vortex or the length of the AF-core soliton depends on not only $\frac{q}{\mu}$, but also $\frac{c_s}{c_n}$.
The $\frac{q}{\mu}$ dependence comes from the tension coefficient, as demonstrated in detail for $\frac{c_s}{c_n}=0.016$ in Ref.~\cite{takeuchi2020quantum}.
According to the phenomenological theory \cite{takeuchi2020quantum},
the distance $d$ between the two spin spots in an elliptic vortex for $\xi_q\gg \xi_s$ is estimated by
\begin{eqnarray}
  \frac{d}{\xi_n}=\frac{\xi_s}{4\xi_n}\left(\sqrt{1+8\pi\frac{\xi_q}{\xi_s}}-1 \right).
  \label{eq:dxin}
\end{eqnarray}
This estimation is in good agreement with the numerical results for small $\frac{q}{\mu}$, as shown by the red arrow in Figs.~\ref{Fig-Evortex}(a) and (b).

\section{Transition from a normal-core vortex}\label{sec-TN}

Here, we describe the continuous phase transition of the vortex core from the N core to other core states.
In general, the critical behavior of a continuous phase transition can be affected by long-wavelength fluctuations of the order parameter beyond the mean-field prediction based on the Ginzburg-Landau (GL) theory.
In our case, the mean-field theory is applicable to the vortex-core transition at zero temperature, because the ordered state is localized in the restricted space of the vortex core,
 where the effect of the fluctuations is less important.
 Here, we investigate the critical behavior of the continuous vortex-core transition within the mean-field approach based on the Bogoliubov theory and GL formalism.

\subsection{Vortex-core transition as the thermodynamic instability}

The phase transition from the N-core vortex to the F-core or AF-core vortex is continuous, as described by the mean-field theory of the GL model \cite{LandauSTATPHYS}.
When $q$ becomes smaller than a critical value $q_{\rm C}$,
the vortex state is a saddle point or the thermodynamic energy $G$ is ``convex upwards'' with respect to a certain fluctuation in the configuration space of $\Psi_m$.
This instability is referred as the thermodynamic instability.
Then, the energy of an elementary excitation or a quasiparticle in the quantum fluid is negative, leading to spontaneous creation and amplification of the excitation to reduce the energy in the dissipative system, which is called the Landau instability in the context of low temperature physics.

The thermodynamic instability is evaluated by the Bogoliubov theory.
The bosonic quasiparticle is described as a collective excitation, the fluctuation of the order parameters $\delta \Psi_m(t,{\bm r})=\Psi_m(t,{\bm r})-\Phi_m({\bm r})$ around the stationary solution $\Phi_m({\bm r})$ of the N-core vortex state.
By linearizing the Lagrangian ${\cal L}(\Phi_m+\delta \Psi_m)$ with respect to $\delta\Psi_m$ and applying the Bogoliubov transformation
\begin{eqnarray}
\delta \Psi_m=u_m({\bm r})e^{-i\omega t}-\left[v_m({\bm r})e^{-i\omega t}\right]^*,
  \label{eq:Btrans}
\end{eqnarray}
one obtains three eigenvalue equations,
\begin{eqnarray}
&& \hbar \omega \binom{u_0}{v_0}=
\begin{pmatrix}
  M_0 & -c_{n}\Phi_0^2 \\
  c_{n}{\Phi_0^*}^2 & -M_0
\end{pmatrix}
\binom{u_0}{v_0},
\label{eq:BdG0}\\
&& \hbar \omega \binom{u_{\pm 1}}{v_{\mp 1}}=
\begin{pmatrix}
  M_\pm & -c_{s}\Phi_0^2 \\
  c_{s}{\Phi_0^*}^2 & -M_{\pm}
\end{pmatrix}
\binom{u_{\pm 1}}{v_{\mp 1}}.
\label{eq:BdGm}
\end{eqnarray}
Here, we used
\begin{eqnarray}
&&
M_{0}=-\frac{\hbar ^{2}}{2M} \nabla ^{2}-\mu +2c_{n}|\Phi_0|^2,
\\
&&
M_{\pm}=-\frac{\hbar ^{2}}{2M} \nabla ^{2}-\mu+q \mp p +(c_{n}+c_{s})|\Phi_0|^2.
\end{eqnarray}
Because the excitation energy is written as
\begin{eqnarray}
\epsilon=\hbar\omega\int d^3x\sum_m(|u_m|^2-|v_m|^2),
\label{eq:Eex}
\end{eqnarray}
the condition for the Landau instability for an excitation is given by $\epsilon <0$.

Equation~(\ref{eq:BdG0}) is independent of the contributions from the $m=\pm 1$ component,
and the eigenvalue problem is identical to that for excitations in an N-core vortex in a scalar BEC.
In this system, there are two excitations with the lowest energy as the Nambu--Goldstone modes with zero energy
 associated with the spontaneous breaking of the U(1) symmetry and translational symmetry, corresponding to the varicose and Kelvin waves \cite{takeuchi2009spontaneous} with zero wave numbers, respectively.
This means that the excitation energy of the nontrivial solutions in Eq.~(\ref{eq:BdG0}) is non-negative.
Therefore, the contributions of $u_0$ and $v_0$ are implicitly neglected in the following discussion to focus on the instability induced by the $m=\pm 1$ component.

The N-core vortex is unstable when the lowest excitation energy
\begin{eqnarray}
     \epsilon_{\rm min}=\min(\epsilon)
\label{eq:Eex_min}
\end{eqnarray}
 becomes negative.
Therefore, the critical point of the vortex-core transition is obtained from the relation
\begin{eqnarray}
  \epsilon_{\rm min}=0.
\label{eq:qc_condition}
\end{eqnarray}

 \subsection{Perturbation analysis for bosonic quasiparticles}
Here, it is revealed that the lowest excitation energy depends on $q$ in a simple manner as follows:
\begin{eqnarray}
  \epsilon_{\rm min}=q-q_{\rm C}.
\label{eq:Emin_qc}
\end{eqnarray}
More concretely, $\frac{q_{\rm C}}{\mu}$ is given by
\begin{eqnarray}
  \frac{q_{\rm C}}{\mu}=-(1+\tilde{M})\frac{c_s}{c_n}-\tilde{\varepsilon}.
  \label{eq:qc_mu}
\end{eqnarray}
with dimensionless constants $\tilde{M}$ and $\tilde{\varepsilon}$.
To obtain these results analytically from Eq.~(\ref{eq:BdGm}),
we extend the method of the perturbation theory for bosonic quasiparticles,
 which was introduced in the context of the splitting instability of a doubly quantized vortex \cite{PhysRevA.63.013602,PhysRevA.74.063620} and succeeded in precisely describing the instability in scalar BECs \cite{takeuchi2018doubly}.
 Here, we extend the theory to the vortex-core instability in spin-1 BECs.

 \subsubsection{Formalism of the perturbation theory}
 For the case of $p=0$, we have $M_+=M_-$ and the eigenvalue equations of $\vec{u}_{\uparrow}=(u_{+1},v_{-1})^{\rm T}$ and $\vec{u}_{\downarrow}=(u_{-1},v_{+1})^{\rm T}$ are identical to each other.
 For simplicity, we drop the suffixes, $\uparrow$ and $\downarrow$, and then the eigenvalue equation is reduced to
 \begin{eqnarray}
\hbar\omega\vec{u}=
\begin{pmatrix}
  \hat{h}+c_s|\Phi_0|^2 & -c_{s}\Phi_0^2 \\
  c_{s}{\Phi_0^*}^2 & -\hat{h}-c_s|\Phi_0|^2
\end{pmatrix}
=(\hat{h}+\delta\hat{h})\vec{u}
\label{eq:BdGper}
 \end{eqnarray}
with $\hat{h}={\rm diag}(h,-h)$,
\begin{eqnarray}
h=-\frac{\hbar ^{2}}{2M} \nabla ^{2}-\mu+q +c_{n}|\Phi_0|^2,
\end{eqnarray}
and
\begin{eqnarray}
\delta\hat{h}=
\begin{pmatrix}
  c_{s}|\Phi_0|^2 & -c_{s}\Phi_0^2 \\
  c_{s}{\Phi_0^*}^2 & -c_{s}|\Phi_0|^2
\end{pmatrix}.
\end{eqnarray}
The mathematical treatment of this problem is similar to the perturbation analysis of single-particle problems in quantum mechanics.
Here, $\hat{h}$ and $\delta\hat{h}$ play the roles of the nonperturbed Hamiltonian operator and the perturbation, respectively.
In contrast, the norm of the wave functions can be positive, negative, and even zero, whereas it is unity (positive) in conventional quantum mechanics.

The eigenvector $\vec{u}$ is represented by a linear combination of a complete set of nonperturbed solutions as follows:
\begin{eqnarray}
   \vec{u}=\sum_{\nu>0} (C_{\nu}\vec{u}_{\nu}+C_{-\nu}\vec{u}_{-\nu}).
   \label{eq:Lcomb}
\end{eqnarray}
Here, $\vec{u}_{\nu}=(\phi_{\nu}, 0)^{\rm T}$, and $\vec{u}_{-\nu}=(0,\phi_{\nu})^{\rm T}$ are the eigenvectors of the eigenequation $\varepsilon_{\pm\nu}\vec{u}_{\pm\nu}=\hat{h}\vec{u}_{\pm\nu}$, reduced to
\begin{eqnarray}
h\phi_{\nu}= \varepsilon_{\nu} \phi_{\nu}
\label{eq:Seq}
\end{eqnarray}
with $\varepsilon_{-\nu}=-\varepsilon_{\nu}$.
The normalization condition
\begin{eqnarray}
\int d^3x \phi_{\nu}^*\phi_{\nu'} =\delta_{\nu,\nu'}
\end{eqnarray}
with the Kronecker delta $\delta_{\nu,\nu'}$
is represented in terms of $\vec{u}_{\pm\nu}$ as
\begin{eqnarray}
{\cal N}_{\pm\nu,\pm\nu'}=\int d^3x \vec{u}_{\pm\nu}^{\dagger}\hat{\sigma}_z\vec{u}_{\pm\nu'}=\pm \delta_{\nu,\nu'},
\end{eqnarray}
with $\sigma_z={\rm diag}(1,-1)$.
According to Eq.~(\ref{eq:Eex}), using this formula,
the nonperturbed excitation energy by the eigenmode of $\pm \nu$ is given by
\begin{eqnarray}
\epsilon=\varepsilon_{\pm \nu}{\cal N}_{\pm\nu,\pm\nu}=\varepsilon_{\nu}.
\label{eq:norm_non}
\end{eqnarray}

\subsubsection{Two-mode approximation}
The Landau instability occurs when there exists at least one excitation with negative energy.
Such an excitation is related to the eigenmode $\vec{u}_{\pm 1}$ with the lowest eigenvalue $\min_\nu(\varepsilon_\nu)=\varepsilon_1$.
As a result, the vortex-core transition is described theoretically by the ground state solution of the single-particle Schr\"{o}dinger problem of Eq.~(\ref{eq:Seq}).
The unperturbed solution is represented by a linear combination of the lowest-eigenvalue solutions in the two-mode approximation as follows:
\begin{eqnarray}
  \vec{u}=C_{+1}\vec{u}_{+1} + C_{-1}\vec{u}_{-1}.
  \label{eq:Ceigen}
\end{eqnarray}
Substituting this formula into Eq.~(\ref{eq:BdGper}),
one obtains the eigenvalue equation
\begin{eqnarray}
  \hbar \omega C_{\pm 1} =\varepsilon_{\pm 1} C_{\pm 1} +M_{\pm 1, \pm 1} C_{\pm 1}+M_{\pm 1, \mp 1} C_{\mp 1}
\end{eqnarray}
with
\begin{eqnarray}
  &&M_{+ 1, + 1}
   = c_s \int d^3x |\Psi_0|^2 |\phi_1|^2=-M_{- 1, - 1}=M_{\rm D},
  \\
  &&M_{+ 1, - 1}
   =-c_s \int d^3x \Psi_0^2 |\phi_1|^2=-M_{- 1, + 1}^*=M_{\rm O}.
\end{eqnarray}
The off-diagonal component $M_{\rm O}$ vanishes as follows.
In the Schr\"{o}dinger problem of Eq.~(\ref{eq:Seq}),
the density profile of the condensate determines the symmetry of the potential for the single-particle wave function $\phi_\nu$.
The ground-state wave function $\phi_1$ is axisymmetric about the $z$ axis, because the condensate density $|\Psi_0|^2=f_0^2$ is axisymmetric about the vortex axis.
Therefore, the integral in $M_{\rm O}$ becomes zero with $\Psi_0=f_0(\rho)e^{i\varphi}$.

From Eq.~(\ref{eq:Ceigen}) with $M_{\rm O}=0$, we have $\hbar\omega= \varepsilon_1+M_{\rm D}$ and $- \varepsilon_1-M_{\rm D}$ for the eigenvectors $(C_{+1},C_{-1})^{\rm T}={\bm C}_+$ and ${\bm C}_-$, respectively.
Here, we used
\begin{eqnarray}
  &&{\bm C}_+=(e^{i\Theta_+},0)^{\rm T},
  \\
  &&{\bm C}_-=(0,e^{i\Theta_-})^{\rm T}
\end{eqnarray}
with constants $\Theta_\pm$.
Substituting these results into Eq.~(\ref{eq:Eex}),
one finds both of the eigensolutions give the same excitation energy,
\begin{eqnarray}
  \epsilon_{\rm min}=\varepsilon_1+M_{\rm D}.
  \label{eq:nonEex}
\end{eqnarray}
When this value is negative, the N-core vortex becomes thermodynamically unstable, and the Landau instability causes the spontaneous creation and condensation of the single-particle state $\phi_1$ in the vortex core.
Because the excitation of $\vec{u}_{\uparrow}$ with the coefficient vector ${\bm C}_{\pm}$ is physically identical to $\vec{u}_{\downarrow}$ with the coefficient ${\bm C}_{\mp}$, we consider only $\vec{u}_{\uparrow}$ in the following.

\subsubsection{Dimensionless constants}
Now, we are ready to compute the dimensionless constants $\tilde{M}$ and $\tilde{\varepsilon}$ in Eq.~(\ref{eq:qc_mu}).
Here, we show that the dimensionless form of Eq.~(\ref{eq:nonEex}) is expressed as
\begin{eqnarray}
  \frac{\epsilon_{\rm min}}{\mu}=\frac{q}{\mu}+(1+\tilde{M})\frac{c_s}{c_n}+\tilde{\varepsilon}
  \label{eq:nd_Eex}
\end{eqnarray}
with the dimensionless constants $\tilde{\varepsilon}\approx -0.25$ and $\tilde{M}\approx 0.45$.

The constants $\tilde{\varepsilon}$ and $\tilde{M}$ are computed by solving the dimensionless version of the Schr\"{o}dinger equation [Eq.~(\ref{eq:Seq})] for $\nu=1$,
\begin{eqnarray}
  \tilde{\varepsilon}\phi_1=-\frac{1}{2}\tilde{\nabla}^2\phi_1+\tilde{V}_{\rm vortex}\phi_1,
  \label{eq:dSch}
\end{eqnarray}
with $\tilde{\varepsilon}=\frac{\varepsilon_1 - q}{\mu}$, $\tilde{\nabla}^2=\xi_n^2\nabla^2$,
  and $\tilde{V}_{\rm vortex}=\tilde{f}_0^2-1$.
Here, $\tilde{f}_0=\sqrt{\frac{c_n}{\mu}|\Phi_0|^2}$ obeys the dimensionless version of Eq.~(\ref{eq:Nvor}),
\begin{eqnarray}
 0=\left[ \frac{1}{2}\left(-\frac{{\rm d}^{2}}{{\rm d}\tilde{\rho}^{2}} -\tilde{\rho}^{-1}\frac{{\rm d}}{{\rm d}\tilde{\rho}}\right)
 -1+\tilde{f}_0^2 \right]\tilde{f}_{0}
\label{eq:dNvor}
\end{eqnarray}
with $\tilde{\rho}=\frac{\rho}{\xi_n}$.
The constant $\tilde{M}$ is computed by
\begin{eqnarray}
 \tilde{M}=\int d^3x \tilde{V}_{\rm vortex}|\phi_1|^2.
\end{eqnarray}
The integral $\xi_n^{-1}\int dz$ is replaced by unity in our case with translational symmetry along the $z$ axis.

\subsubsection{Spin fluctuation}

The distribution of the spin density in the vortex core is reproduced as a result of the condensation of the excitations as follows.
The condensation of the single-particle state $\phi_1$ causes magnetization in the vortex core, because the negative-energy excitations cause a fluctuation in the spin density
whereas the N-core vortex has no magnetization.

The spin fluctuation $\delta s_{x,y,z}$ of $s_{x,y,z}$ by the excitation $\vec{u}=\vec{u}_{\uparrow}$ of ${\bm C}_{\pm}$ is written as
\begin{eqnarray}
  &&\delta s_x = \pm \sqrt{2}\phi_1{\rm Re}(e^{i\Theta_{\pm}}\Phi_0^*)=\pm \sqrt{2}f_0\phi_1\cos(\varphi+\Theta_{\pm}),
  \\
  &&\delta s_y = \sqrt{2}\phi_1{\rm Im}(\Phi_0e^{-i\Theta_{\pm}})=\sqrt{2}f_0\phi_1\sin(\varphi-\Theta_{\pm}),
  \\
  &&\delta s_z = \pm\phi_1^2.
\end{eqnarray}
Here, $\phi_1$ and $f_0$ are assumed to be positive functions without loss of generality.

The condensation of the excitation of ${\bm C}_{\pm}$ leads to an F-core vortex because each excitation has a finite magnetization along the $z$ axis.
The wave function $\phi_1$ is localized as a bound state in the vortex core and
the ${\bm C}_+$ excitation with $\Theta_+=0$ reproduces the localized Mermin--Ho texture like the right panel in Fig.~\ref{Fig-Cross}(b).
As observed in the F-core vortex with $q\sim q_{\rm C}$ in Fig.~\ref{Fig-FcoreV}(c),
the amplitude of $s_z=\delta s_z$ is small compared to $s_\bot=\sqrt{\delta s_x^2+\delta s_y^2}$, because the former and the latter are proportional to $\phi_1$ and $\phi_1^2(\ll \phi_1)$, respectively.

The transition to the AF-core vortex is induced by the combination of the excitations of ${\bm C}_+$ and ${\bm C}_-$.
The sum of the spin fluctuations with $\Theta_+=\Theta_-$ causes $s_y\neq 0$ and $s_x=s_z=0$.
The distribution of the spin density is similar to that of the right panel in Fig.~\ref{Fig-Cross}(a).
 The direction of the spin density on the $xy$ plane is arbitrarily determined by changing $\Theta_\pm$ with $\Theta_+-\Theta_-=0$ or $\pi$.

 Note that the excitations that cause the transition to the AF- and F-core vortices have the same energy in the perturbation theory.
 Therefore, the above analysis does not tell us which of the AF- and F-core vortices is the lowest-energy vortex.
 This is because the second-order term ($\propto \delta s_z^2$) associated with the spin interaction is neglected in the Bogoliubov formalism. This inconsistency may be resolved by taking into account the next-to-leading-order terms or corrections.
 A qualitative perspective was presented to explain the discontinuous phase transition between the AF- and F-core vortices at $c_s=0$ in Sec.~\ref{subsec-CP}, and thus we cease to consider the higher-order corrections.

\subsubsection{Comparisons with the full Bogoliubov theory}

 To examine the validity of the result of the perturbation analysis,
 we compare it with the numerical result of the full eigenvalue problem.
The full eigenvalue equations, as obtained by linearizing the Lagrangian with respect to $u_m$ and $v_m$, were diagonalized numerically.
The equation reduces to the one-dimensional equation of $[\bar{u}_m, \bar{v}_m]$ by writing as $[u_m({\bm r}),v_m({\bm r})]=[\bar{u}_m(\rho)e^{iL_m\varphi},\bar{v}_m(\rho)e^{-iL_m\varphi}]e^{il\varphi}$ with $L_m=1$
\footnote{The full eigenvalue equation of $[\bar{u}_m, \bar{v}_m]$ was numerically diagonalized by using the Intel$\textsuperscript{\textregistered}$ Math Kernel Library LAPACK with a grid size $\Delta \rho=0.125\xi_n$.}
.
The lowest excitation energy $\epsilon_{\rm min}$ appears in the energy spectrum of $l=-1$,
 where the centrifugal potential ($\propto \frac{L_m+l}{\rho^2}$) vanishes for the wave function $\bar{u}_m$.

\begin{figure}
\begin{center}
\includegraphics[width=1.0 \linewidth, keepaspectratio]{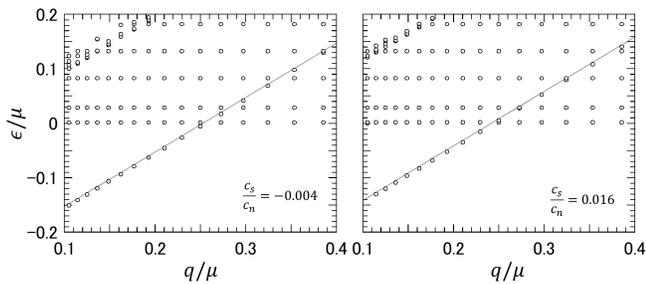}
\end{center}
\vspace{-5mm}
\caption{
Excitation spectrum of $l=-1$ around the critical point for $\frac{c_s}{c_n}=-0.004$ (left) and $0.016$ (right).
The plots show the lowest values of excitation energy. The solid lines represent Eq.~(\ref{eq:Emin_qc}) with the critical value of Eq.~(\ref{eq:qc_mu}).
}
\label{Fig-BdG}
\end{figure}

The numerical result of the excitation spectrum of an N-core vortex with $L_m=1$ is plotted together with the result of the perturbation theory in Fig.~\ref{Fig-BdG}.
The perturbation theory explains well the numerical result of the negative-energy modes, which occur below the critical point $q_{\rm C}$.
The zero-energy Kelvin mode ($l=-1$) corresponds to the data distributed along the $\epsilon=0$ line while the varicose mode with $l=0$ does not appear in this plot.
We also observe the linear $q$ dependence of the lowest-energy spectrum even for larger values of $\frac{c_s}{c_n}$,
whereas a slightly greater difference is found between the theoretical result and the numerical result (not shown).
Considering that spin-1 BECs are realized mostly for very small $\frac{|c_s|}{c_n}$,
the difference is sufficiently small and the theoretical prediction is available on a practical level for making the phase diagram of Fig.~\ref{Fig-diagram}.
The smallness of the difference is presented in a later discussion on the critical behavior in Sec.~\ref{sec:CB} (see Fig.~\ref{Fig-Qc}).

 \subsection{Critical behavior}\label{sec:CB}
 The critical behavior of the continuous vortex-core phase transition is investigated.
 In the transition from an N-core vortex to an F- or AF-core vortex,
 the condensate density grows from zero at the vortex axis.
 Accordingly, we regard $\Psi_{\pm 1}(0)$ at the axis as the order parameter of the continuous transition and evaluate the critical behaviors of $n(0)=|\Psi_{+ 1}(0)|^2+|\Psi_{- 1}(0)|^2$ and $s_z(0)=|\Psi_{+ 1}(0)|^2 - |\Psi_{- 1}(0)|^2$.

The phenomenology of the GL model is applied to the vortex-core transition.
 The vortex energy $E_{\rm vortex}$ is phenomenologically represented by a power-series expansion of the effective order parameter $\Psi_{\pm 1}(0)$.
 By taking into account the form of the energy functional $G$, the energy is represented as
\begin{eqnarray}
E_{\rm vortex}=E_{\rm N}+ \alpha (q-q_{\rm C}) n(0)+\beta_nn(0)^2+\beta_ss_z(0),
\end{eqnarray}
where $E_{\rm N}$ is the vortex energy of an N-core vortex, and the spatial variation along the $z$ axis is neglected.
The vortex core is in an ordered state for $q<q_{\rm C}$, and the energy is minimized with
\begin{eqnarray}
n(0)=\frac{\alpha}{2\beta_n} (q_{\rm C}-q)~~~(q\to q_{\rm C}).
\label{eq:GL}
\end{eqnarray}
The sign of $\beta_s$ is derived from that of $c_s$ and we have
$s_z=0$ for $\beta_s>0$ and $s_z= \pm n(0)$ for $\beta_s<0$,
 representing the AF-core and F-core vortices, respectively.

In the hydrostatic approximation with $\Psi_0=0$ and $p=0$, the core density is given by $n(0)=n_{\rm AF}=\frac{\mu-q}{c_n}$ for the AF-core vortex and $n(0)=n_{\rm F}=\frac{\mu-q/2}{c_n+c_s}$ for the F-core vortex.
This approximation is valid for small $q$, such that the spatial gradient of the order parameters is negligibly small with a large vortex-core size.
Considering the interpolation between the hydrostatic approximation and the GL model,
we evaluate the core density by
\begin{eqnarray}
n(0)=n_{\rm eq}\left(1-\frac{q}{q_{\rm C}}\right)
\label{eq:intF}
\end{eqnarray}
with $n_{\rm eq}=n_{\rm AF}$ for $c_s>0$ and $n_{\rm eq}=n_{\rm F}$ for $c_s<0$.

\begin{figure}
\begin{center}
\includegraphics[width=1.0 \linewidth, keepaspectratio]{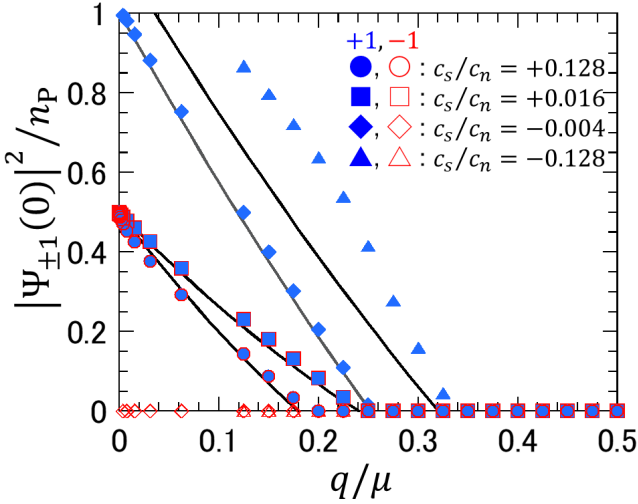}
\end{center}
\vspace{-5mm}
\caption{
The critical behavior of the condensate density $|\Psi_{\pm 1}(0)|^2$ at the vortex axis (${\bm \rho}=0$) in the AF- and F-core vortices for $\frac{c_s}{c_n}=-0.128$, $-0.004$, $0.016$, and $0.128$.
The solid and open symbols represent $|\Psi_{+ 1}(0)|^2$ and $|\Psi_{- 1}(0)|^2$, respectively.
The $m=-1$ component vanishes at the vortex axis in the F-core vortex with $L=-N=1$.
The solid curves show the results obtained using Eq.~(\ref{eq:intF}), with the critical value of Eq.~(\ref{eq:qc_mu}).
}
\label{Fig-Qc}
\end{figure}

In Fig.~\ref{Fig-Qc}, the numerical result of the core density $n(0)$ is plotted as a function of $\frac{q}{\mu}$, together with the interpolating formula of Eq.~(\ref{eq:intF}).
The core density is nearly a linear function of $\frac{q}{\mu}$, which is consistent with the critical behavior of Eq.~(\ref{eq:GL}).
For $c_s>0$, both the $m=+1$ and $m=-1$ components grow from zero with the same amplitude, whereas the amplitude of the $m=\pm 1$ component remains zero in the F-core vortex with $s_z=\mp n(0)$.
The critical value $q_{\rm C}$ is simply estimated from the numerical data by applying the least-squares method to the three smallest values of $\frac{|\Psi_{\pm 1}(0)|^2}{n_{\rm P}} (>10^{-7})$, and we obtain $\frac{q_{\rm C}}{\mu}=0.3345$, $0.2514$, $0.2430$, and $0.1902$ for $\frac{c_s}{c_n}=-0.128$, $-0.004$, $0.016$, and $0.128$, respectively.
These estimations are in good agreement with the prediction of the perturbation analysis with a small error,
 which gives $\frac{q_{\rm C}}{\mu}\approx 0.32$, $0.25$, $0.24$, and $0.17$.
It is found that Eq.~(\ref{eq:intF}) agrees with the numerical data quantitatively, not only near the critical point but also in all parameter regimes, except for the case of $\frac{c_s}{c_n}=-0.128$.
This inconsistency indicates the lack of validity of the hydrostatic approximation for large negative $\frac{c_s}{c_n}$;
we have no solution of the F-core vortex for $\frac{q}{\mu}<-2\frac{c_s}{c_n}$, where the bulk is not the P phase but the BA phase.
We also found that the transverse spin density $s_\bot$ is no longer localized around the vortex axis and is broadly distributed with a finite amplitude for $\frac{q}{\mu}\geq 0.25$ in the numerical solution of $\frac{c_s}{c_n}=-0.128$.
This effect may be an additional reason for the inconsistency, and is discussed as the finite-size effect in Sec.~\ref{sec:FSE}.

\section{Finite-size effect}\label{sec:FSE}
The effect of trapping potentials can be crucial in actual experiments for small $\frac{q}{\mu}$, for which
 the size of the vortex core becomes comparable to the size of the atomic cloud.
Most previous studies on vortices in spinor BECs considered vortices in trapped systems, but did not take into account the quadratic Zeeman effect \cite{isoshima2001quantum}.
Here, we briefly mention the effect of the trapping potential on the vortices based on the above theoretical analyses.

First, we consider the finite-size effect in a cylindrical box with a radius of $R$.
In a spin-1 BEC with ferromagnetic interaction of $c_s<0$,
the bulk P phase disappears if the outer radius $\rho_{\rm BA}$ of the BA skin is larger than $R$, $\rho_{\rm BA}>R$.
In such a small system,
the $m=\pm 1$ components are not localized around the vortex axis but are distributed broadly.
This is just the vortex solution obtained numerically for small $\frac{q}{\mu}$ with $\frac{c_s}{c_n}=-0.128$, as mentioned in the last paragraph in Sec.~\ref{sec:CB}.
The vortex is regarded as an F-core vortex in the BA phase in the sense that the ordered state outside the F core is in the local BA phase within the hydrostatic approximation.
For $c_s>0$, it is found that the vortex structure disappears when the width $d$ of an elliptic vortex becomes comparable to the system size for small $\frac{q}{\mu}$.
This effect implies the difficulty of preparing vortices in a trapping system for $q\to 0$.
This is similar to the situation in the AF phase, where the vortex-core size ($\sim \xi_q$) of a nematic-spin vortex diverges for small negative $q$ \cite{underwood2020properties}.

Next, we consider the effect of nonuniformity due to a trapping potential.
For simplicity, we assume the same external potential $V_m({\bm r})=V_{\rm ext}({\bm r})$ for all components.
In a similar manner to the formalism in the hydrostatic approximation,
 this effect is included effectively as the local chemical potential
 \begin{eqnarray}
 \mu_{\rm ext}({\bm r})=\mu-V_{\rm ext}({\bm r}).
 \label{eq:mu_trap}
 \end{eqnarray}
The density healing length also varies spatially according to Eq.~(\ref{eq:xi_n}) by replacing $\mu \to \mu_{\rm ext}$.
This replacement deforms the phase diagram in Fig.~\ref{Fig-PDbulk} because the parameters of the vertical and horizontal axes are also shited.
Accordingly, the critical point $q_{\rm C}$ of the vortex-core transition is shifted as $q_{\rm C}\to q_{\rm C}^{\rm local} = q_{\rm C} \frac{\mu_{\rm ext}}{\mu}$.
Therefore, in a spinor condensate trapped by a harmonic potential,
the local critical value $q_{\rm C}^{\rm local}$ in the trap center with $V_{\rm ext}=0$ is smaller than that of the outer region with $V_{\rm ext}>0$.
This implies that the core structure of a vortex can change when it moves away from the center of the trap.
It is not difficult to take into account the effect due to a magnetic trap.
 For example, the potential $V_m$ depends on $m$, and the local Zeeman shift induced by a spatial gradient of the magnetic field can be included as the local shift of $p$ and/or $q$, as the centrifugal potential is given by Eqs.~(\ref{eq:qb}) and (\ref{eq:pb}).

\section{Summary and discussion} \label{sec-S}
The phase diagram of vortices in the P phase of spin-1 BECs is theoretically obtained by revealing the parameter dependence of the vortex-core structure in a singly quantized vortex.
We found three types of vortices depending on the dimensionless parameters $\frac{q}{\mu}$ and $\frac{c_s}{c_n}$, the F-core, AF-core, and N-core vortices, as the lowest-energy state of a singly quantized vortex.
The N-core vortex is identical to the conventional vortex in scalar BECs and is the lowest-energy vortex when the quadratic Zeeman coefficient $q$ exceeds a critical value $q_{\rm C}$, as given by Eq.~(\ref{eq:qc_mu}).
The perturbation theory of bosonic quasiparticles reveals that the continuous vortex-core transition of an N-core vortex is induced by the Landau (thermodynamic) instability.
The vortex solution with the lowest energy below $q_{\rm C}$ is
 the axisymmetric F-core vortex with a BA skin for ferromagnetic interaction ($c_s<0$), whereas it is the elliptic AF-core vortex with BA edges for $c_s>0$.
The AF- and F-core vortices are energetically degenerate at the critical point $c_s=0$ of the discontinuous phase transition, and the former and the latter can be metastable states for $c_s<0$ and $c_s>0$, respectively.
In fact, in the later stage of the quench dynamics \cite{kangPhysRevLett.122.095301,kangPhysRevA.101.023613}, a few vortices were found to possess longitudinal spin density in the vortex core, corresponding to the metastable F-core vortices.

The hydrostatic approximation based on the vortex winding rule is useful for qualitatively describing the vortex-core structure and explains the localized ferromagnetic-spin texture in the F-core vortex, whose size is characterized by the Zeeman length $\xi_q$.
The influence of the trapping potential is also discussed, and becomes more important in the experiments with smaller $\frac{q}{\mu}$ because the core size of the vortices diverges as $\xi_q\propto q^{-1/2}$ for $q \to 0$.
These predictions can be examined experimentally, because the considered parameter regimes cover those of typical experiments, such as spin-1 condensates of $^{23}$Na with small positive $\frac{c_s}{c_n}$ and those of $^{87}$Rb and $^{41}$K with small negative $\frac{c_s}{c_n}$,
 whereas the parameter $\frac{q}{\mu}$ is well controlled experimentally \cite{PhysRevA.73.041602,PhysRevA.89.023608}.
 The hydrostatic approximation would be useful for understanding the vortex-core structure in the F and BA phases and the impact of the finite-size effect in the recent experiment \cite{weiss2019controlled} as was analyzed in Sec.~\ref{sec:FSE}.

The dynamics of topological defects in multicomponent superfluids is a challenging problem to provide a rich variety of physical phenomena because of the multiple degrees of freedom.
It is important to understand the local ordered states and their distribution in the core of topological defects,
because the dynamics may depend on them not only quantitatively but also qualitatively.
The core transition of solitons in spinor BECs is a timely topic.
For example, solitons in the P phase of spin-1 BECs are classified as AF-core, BA-core, F-core, and N-core solitons according to Ref.~\cite{liu2020phase}.
Very recently, it has been observed that the core transition occurs in the collision of solitons in experiments of a spin-1 BEC in the polar phase. They found that the two solitons after the collision possess finite longitudinal spin density, whereas they do not before the collision \cite{PhysRevLett.125.170401}.
The magnetized solitons correspond to the F-core solitons in the phase diagram of solitons predicted theoretically in Ref.~\cite{liu2020phase}.
Desirably, the transverse spin density is observed in addition to the longitudinal one,
because the initial solitons can be identified as AF-core or BA-core solitons according to the nematic-spin order and transverse spin density.
Therefore, precise measurement of the core structure of topological defect is important for understanding the dynamics of topological defects in multicomponent superfluids on a fundamental level.

 We clarify the difference between our study and the preceding one in Ref.~\cite{PhysRevA.93.033633} in order to avoid confusion.
 The vortex state with localized texture in the right-hand panel of Fig.~\ref{Fig-Cross}(b) looks similar to those obtained in Fig.~4 (left) of Ref.~\cite{PhysRevA.93.033633} (see also Fig.~1 (top left) of Ref.~\cite{PhysRevLett.112.075301}).
 The latter was realized in a parameter regime called the ``polar regime'' ($c_s>0$)
 whereas the former is realized for $-\frac{q}{2\mu}<\frac{c_s}{c_n}<0$ and $q>0$ with $p=0$.
Although they did not specify the values of $p$ and $q$ in the numerical simulations based on an unusual thermodynamic treatment,
it seems to be realized for $q<0$ under the effective ``bias'' of nonzero $p$ caused by the numerical renormalization of the longitudinal magnetization \cite{Magnus_PC}.
The spin distribution in the right-hand panel of Fig.~\ref{Fig-Cross}(a) is similar to that in the vortex state obtained in Fig.~8 of the same paper, where the values of $p$ and $q$ were not specified again (see also Fig.~1(a) in Ref.~\cite{PhysRevLett.116.085301}).
As was pointed out in Ref.~\cite{takeuchi2020quantum},
they did not consider the impact of the domain wall (the AF-core soliton) jointing the two spin spots regardless of the values of $p$ and $q$,
and could not distinguish essentially the vortex state in the P phase from a pair of half quantum vortices in the AF phase under external rotation \cite{PhysRevA.86.013613}.

It is also fruitful to connect the theory and formulation with the bound state in the core of topological defects in other superfluids,
 including fermionic superfluids and superconductors beyond the GP and Bardeen--Cooper--Schrieffer (BCS) models.
The condensates of the bound state of boson pairs or multiple fermions can occur in a quantized vortex \cite{salomaa1989exotic},
and even lead to the nonaxisymmetric vortex in scalar superfluids \cite{volovik2002half}.
The vortex winding rule can be connected with the symmetry of such exotic condensates,
 and then to explore the possibility of nonaxisymmetric vortices as a result of the local condensation would be an interesting future prospect of this work.

\begin{acknowledgments}
H.T. thanks G. E. Volovik for suggesting the topics of the condensation of boson pairs and multiple fermions in the vortex core.
This work was supported by JSPS KAKENHI Grants No. JP18KK0391, No. JP20H01842, and No. 20H01843,
 and in part by the OCU ``Think globally, act locally'' Research Grant for Young Scientists 2020 through the hometown donation fund of Osaka City.
\end{acknowledgments}


\begin{thebibliography}{10}

\bibitem{vollhardt2013superfluid}
Dieter Vollhardt and Peter Wolfle.
\newblock {\em The superfluid phases of helium 3}.
\newblock Courier Corporation, North Chelmsford, 2013.

\bibitem{volovik2003universe}
Grigory~E Volovik.
\newblock {\em The universe in a helium droplet}, volume 117.
\newblock Oxford University Press on Demand, New York, 2003.

\bibitem{kasamatsu2005vortices}
Kenichi Kasamatsu, Makoto Tsubota, and Masahito Ueda.
\newblock {Vortices in multicomponent Bose--Einstein condensates}.
\newblock {\em International Journal of Modern Physics B}, 19(11):1835--1904,
  2005.

\bibitem{kawaguchi2012spinor}
Yuki Kawaguchi and Masahito Ueda.
\newblock {Spinor Bose--Einstein Condensates}.
\newblock {\em Physics Reports}, 520(5):253--381, 2012.

\bibitem{takeuchi2020quantum}
Hiromitsu Takeuchi.
\newblock {Quantum elliptic vortex in a nematic-spin Bose-Einstein condensate}.
\newblock {\em arXiv preprint arXiv:2009.03556}, 2020.

\bibitem{kangPhysRevLett.122.095301}
Seji Kang, Sang~Won Seo, Hiromitsu Takeuchi, and Y.~Shin.
\newblock {Observation of Wall-Vortex Composite Defects in a Spinor
  Bose-Einstein Condensate}.
\newblock {\em Phys. Rev. Lett.}, 122:095301, Mar 2019.

\bibitem{liu2020phase}
I-Kang Liu, Shih-Chuan Gou, and Hiromitsu Takeuchi.
\newblock {Phase diagram of solitons in the polar phase of a spin-1
  Bose-Einstein condensate}.
\newblock {\em Phys. Rev. Research}, 2:033506, Sep 2020.

\bibitem{underwood2020properties}
Andrew P.~C. Underwood, D.~Baillie, P.~Blair Blakie, and H.~Takeuchi.
\newblock {Properties of a nematic spin vortex in an antiferromagnetic spin-1
  Bose-Einstein condensate}.
\newblock {\em Phys. Rev. A}, 102:023326, Aug 2020.

\bibitem{isoshima2001quantum}
Tomoya Isoshima, Kazushige Machida, and Tetsuo Ohmi.
\newblock {Quantum vortex in a spinor Bose-Einstein condensate}.
\newblock {\em Journal of the Physical Society of Japan}, 70(6):1604--1610,
  2001.

\bibitem{PhysRevA.93.033633}
Justin Lovegrove, Magnus~O. Borgh, and Janne Ruostekoski.
\newblock {Stability and internal structure of vortices in spin-1 Bose-Einstein
  condensates with conserved magnetization}.
\newblock {\em Phys. Rev. A}, 93:033633, Mar 2016.

\bibitem{katsimiga2020phase}
GC~Katsimiga, SI~Mistakidis, P~Schmelcher, and PG~Kevrekidis.
\newblock {Phase Diagram, Stability and Magnetic Properties of Nonlinear
  Excitations in Spinor Bose-Einstein Condensates}.
\newblock {\em arXiv preprint arXiv:2008.00475}, 2020.

\bibitem{PhysRevLett.125.170401}
Stefan Lannig, Christian-Marcel Schmied, Maximilian Pr\"ufer, Philipp Kunkel,
  Robin Strohmaier, Helmut Strobel, Thomas Gasenzer, Panayotis~G. Kevrekidis,
  and Markus~K. Oberthaler.
\newblock {Collisions of Three-Component Vector Solitons in Bose-Einstein
  Condensates}.
\newblock {\em Phys. Rev. Lett.}, 125:170401, Oct 2020.

\bibitem{PhysRevLett.103.080603}
Ari~M. Turner.
\newblock {Mass of a Spin Vortex in a Bose-Einstein Condensate}.
\newblock {\em Phys. Rev. Lett.}, 103:080603, Aug 2009.

\bibitem{PhysRevA.94.063615}
Lewis~A. Williamson and P.~B. Blakie.
\newblock {Dynamics of polar-core spin vortices in a ferromagnetic spin-1
  Bose-Einstein condensate}.
\newblock {\em Phys. Rev. A}, 94:063615, Dec 2016.

\bibitem{williamson2020damped}
Lewis~A Williamson and PB~Blakie.
\newblock {A damped point-vortex model for polar-core spin vortices in a
  ferromagnetic spin-1 Bose-Einstein condensate}.
\newblock {\em arXiv preprint arXiv:2010.12154}, 2020.

\bibitem{PhysRevLett.75.3320}
\"U. Parts, J.~M. Karim\"aki, J.~H. Koivuniemi, M.~Krusius, V.~M.~H. Ruutu,
  E.~V. Thuneberg, and G.~E. Volovik.
\newblock {Phase Diagram of Vortices in Superfluid ${}^{3}$
  $He\ensuremath{-}\mathit{A}$}.
\newblock {\em Phys. Rev. Lett.}, 75:3320--3323, Oct 1995.

\bibitem{PhysRevB.101.024517}
Robert~C. Regan, J.~J. Wiman, and J.~A. Sauls.
\newblock {Vortex phase diagram of rotating superfluid
  $^{3}\mathrm{He}\text{\ensuremath{-}}\mathrm{B}$}.
\newblock {\em Phys. Rev. B}, 101:024517, Jan 2020.

\bibitem{ohmi1998bose}
Tetsuo Ohmi and Kazushige Machida.
\newblock {Bose-Einstein condensation with internal degrees of freedom in
  alkali atom gases}.
\newblock {\em Journal of the Physical Society of Japan}, 67(6):1822--1825,
  1998.

\bibitem{kangPhysRevA.101.023613}
Seji Kang, Deokhwa Hong, Joon~Hyun Kim, and Y.~Shin.
\newblock {Crossover from weak to strong quench in a spinor Bose-Einstein
  condensate}.
\newblock {\em Phys. Rev. A}, 101:023613, Feb 2020.

\bibitem{chandrasekhar_1992}
S.~Chandrasekhar.
\newblock {\em Liquid Crystals}.
\newblock Cambridge University Press, 2 edition, 1992.

\bibitem{donnelly1991quantized}
Russell~J Donnelly.
\newblock {\em Quantized vortices in helium II}, volume~2.
\newblock Cambridge University Press, 1991.

\bibitem{tsubota2013quantum}
Makoto Tsubota, Michikazu Kobayashi, and Hiromitsu Takeuchi.
\newblock {Quantum hydrodynamics}.
\newblock {\em Physics Reports}, 522(3):191--238, 2013.

\bibitem{Note1}
The steepest descent method is employed to obtain the vortex solution
  numerically. The radial coordinate $\rho $ is discretized as $\rho \to \rho
  _i=i\Delta \rho ~(i=0,1,2,\protect \cdots )$ with a grid size $\Delta \rho
  =0.25\xi _n$. The radial derivatives of $f_m$ are computed with the finite
  difference approximation; for example, $\protect \frac {{\protect \rm
  d}}{{\protect \rm d}\rho }f_m$ and $\protect \frac {{\protect \rm
  d}^2}{{\protect \rm d}\rho ^2}f_m$ are computed by the central difference of
  the first and second orders, respectively.

\bibitem{LandauSTATPHYS}
L.~D. Landau and E.~M. Lifshitz.
\newblock {Statistical Physics, Part 1}.
\newblock {\em (Pergamon Press, New York, 1980)}.

\bibitem{takeuchi2009spontaneous}
Hiromitsu Takeuchi, Kenichi Kasamatsu, and Makoto Tsubota.
\newblock {Spontaneous radiation and amplification of Kelvin waves on quantized
  vortices in Bose-Einstein condensates}.
\newblock {\em Physical Review A}, 79(3):033619, 2009.

\bibitem{PhysRevA.63.013602}
Dmitry~V. Skryabin.
\newblock {Instabilities of vortices in a binary mixture of trapped
  Bose-Einstein condensates: Role of collective excitations with positive and
  negative energies}.
\newblock {\em Phys. Rev. A}, 63:013602, Dec 2000.

\bibitem{PhysRevA.74.063620}
Emil Lundh and Halvor~M. Nilsen.
\newblock {Dynamic stability of a doubly quantized vortex in a
  three-dimensional condensate}.
\newblock {\em Phys. Rev. A}, 74:063620, Dec 2006.

\bibitem{takeuchi2018doubly}
Hiromitsu Takeuchi, Michikazu Kobayashi, and Kenichi Kasamatsu.
\newblock {Is a doubly quantized vortex dynamically unstable in uniform
  superfluids?}
\newblock {\em Journal of the Physical Society of Japan}, 87(2):023601, 2018.

\bibitem{Note2}
The full eigenvalue equation of $[\protect \bar {u}_m, \protect \bar {v}_m]$
  was numerically diagonalized by using the Intel$\protect \textsuperscript
  {\textregistered }$ Math Kernel Library LAPACK with a grid size $\Delta \rho
  =0.125\xi _n$.

\bibitem{PhysRevA.73.041602}
Fabrice Gerbier, Artur Widera, Simon F\"olling, Olaf Mandel, and Immanuel
  Bloch.
\newblock {Resonant control of spin dynamics in ultracold quantum gases by
  microwave dressing}.
\newblock {\em Phys. Rev. A}, 73:041602, Apr 2006.

\bibitem{PhysRevA.89.023608}
L.~Zhao, J.~Jiang, T.~Tang, M.~Webb, and Y.~Liu.
\newblock {Dynamics in spinor condensates tuned by a microwave dressing field}.
\newblock {\em Phys. Rev. A}, 89:023608, Feb 2014.

\bibitem{weiss2019controlled}
Lauren~S Weiss, Magnus~O Borgh, Alina Blinova, Tuomas Ollikainen, Mikko
  M{\"o}tt{\"o}nen, Janne Ruostekoski, and David~S Hall.
\newblock {Controlled creation of a singular spinor vortex by circumventing the
  Dirac belt trick}.
\newblock {\em Nature communications}, 10(1):1--8, 2019.

\bibitem{PhysRevLett.112.075301}
Justin Lovegrove, Magnus~O. Borgh, and Janne Ruostekoski.
\newblock {Energetic Stability of Coreless Vortices in Spin-1 Bose-Einstein
  Condensates with Conserved Magnetization}.
\newblock {\em Phys. Rev. Lett.}, 112:075301, Feb 2014.

\bibitem{Magnus_PC}
{Magnus O. Borgh (private communication) }.

\bibitem{PhysRevLett.116.085301}
Magnus~O. Borgh, Muneto Nitta, and Janne Ruostekoski.
\newblock Stable core symmetries and confined textures for a vortex line in a
  spinor bose-einstein condensate.
\newblock {\em Phys. Rev. Lett.}, 116:085301, Feb 2016.

\bibitem{PhysRevA.86.013613}
Justin Lovegrove, Magnus~O. Borgh, and Janne Ruostekoski.
\newblock Energetically stable singular vortex cores in an atomic spin-1
  bose-einstein condensate.
\newblock {\em Phys. Rev. A}, 86:013613, Jul 2012.

\bibitem{salomaa1989exotic}
MM~Salomaa and GE~Volovik.
\newblock {Exotic states in the cores of quantised vortices for superfluids and
  superconductors}.
\newblock {\em Journal of Physics: Condensed Matter}, 1(1):277, 1989.

\bibitem{volovik2002half}
GE~Volovik.
\newblock {Half-quantum vortices in strongly correlated Bose liquids}.
\newblock {\em arXiv preprint cond-mat/0208555}, 2002.

\end{thebibliography}


\end{document}